\documentclass{article}
\usepackage{epsfig}

\begin{document}

\title{Fitting of $Z'$ parameters}

\author{Alexey Gulov\footnote{gulov@dsu.dp.ua}~ and Vladimir
Skalozub\footnote{skalozubv@daad-alumni.de}\\
{\it Dnipropetrovsk National University, Dnipropetrovsk, Ukraine}}

\maketitle

\begin{abstract}
The paper deals with an approach to the model-independent searching
for the $Z'$ gauge boson as a virtual state in scattering
processes. The relations between the $Z'$ couplings to fermions
covering a wide class of models beyond the standard model are
found and used. They reduce in an essential way the number of
parameters to be fitted in experiments. Special observables
which uniquely pick out the $Z'$ at energies of LEP and ILC colliders
in different leptonic processes
are introduced and the data of LEP experiments are analyzed. The
$Z'$ couplings to leptons and quarks are estimated at 95\%
confidence level. At this level, the LEP data are compatible with
the existence of the $Z'$ with the mass $m_{Z'} \sim 1 - 1.2$ TeV.
These estimates may serve as a guide for experiments at the
Tevatron and/or LHC. A comparison with other approaches and
results is given.
\end{abstract}

\section{Introduction}

The precision test of the standard model (SM) at the LEP gave a
possibility not only to determine all the parameters and particle
masses at the level of radiative corrections but also afforded an
opportunity for searching for signals of new heavy particles
beyond the energy scale of it. On the base of the LEP2 experiments
the low bounds on parameters of various models extending the SM
have been estimated and the scale of new physics was obtained
\cite{Alcaraz:2006mx,Abbiendi:2003dh,Abbiendi:1998ea,Ackerstaff:1997nf,Abdallah:2005ph}.
Although no new particles were discovered, a general believe is
that the energy scale of new physics to be of order 1~TeV, that
may serve as a guide for experiments at the Tevatron and  LHC. In
this situation, any information about new heavy particles obtained
on the base of the present day data is desirable and important.

Numerous extended models include the $Z'$ gauge boson -- massive
neutral vector particle associated with the extra $U(1)$ subgroup
of an underlying group. Searching for this particle as a virtual
state is widely discussed in the literature (see Ref.
\cite{Leike:1998wr,Langacker:2008yv} for review). In the content
of searching for $Z'$ at the LHC and the ILC an essential
information and prospects for future investigations are given in
lectures \cite{Rizzo:2006nw}. Such aspects as the mass of $Z'$,
couplings to the SM particles, $Z$--$Z'$ mixing and its influence
in various processes and particles parameters, distinctions
between different models are discussed in details. We shall turn
to these papers in what follows. As concerned a searching for $Z'$
in the LEP experiments and the experiments at Tevatron
\cite{Ferroglia:2006mj}, it was carried out mainly in a
model-dependent way. Some popular models has been investigated and
low bounds on the mass $m_{Z'}$ were estimated (see Refs.
\cite{Alcaraz:2006mx,Abbiendi:2003dh,Abbiendi:1998ea,Ackerstaff:1997nf,Abdallah:2005ph},
the recent results in Refs. \cite{Erler:2009jh,aguila}). As it is
occurred, the low masses are varying in a wide energy interval
400-1800 GeV dependently on a specific model. These bounds are a
little bit different in the LEP and Tevatron experiments. In this
situation a model-independent analysis is of interest.

In the papers
\cite{Gulov:2000eh,Gulov:1999ry,Demchik:2003pd,Gulov:2004sg} of
the present authors a new approach for the model-independent
search for $Z'$-boson was proposed. In contrast to other
model-independent searches, it gives a possibility to pick out
uniquely $Z'$ virtual state proper  for a wide class of models
(listed below) beyond the SM. Our consideration is based on two
constituents: 1) The relations between $Z'$ couplings motivated by
renormalizability of an unknown theory beyond the SM. Due to these
relations, a number of unknown $Z'$ parameters entering the
amplitudes of different scattering processes considerably
decreases. 2) When these relations are accounted for, some
kinematics properties of the amplitude become uniquely correlated
with this virtual state and the $Z'$ signals exhibit themselves.
The corresponding observables have also been introduced and
applied to analyze the LEP2 experiment data. Comparing the mean
values of the observables with the necessary specific values, one
could arrive at a conclusion about the $Z'$ existence. The
confidence level (CL) of these values has been estimated and
adduced in addition. Without taking into consideration the
relations between coupling the determination of $Z'$-boson
requires a supplementary specification due to a larger number of
different couplings contributing to the observables.

In Refs. \cite{Gulov:2000eh,Demchik:2003pd,Gulov:2004sg} the
one-parametric observables were introduced and the signals (hints
in fact) of the $Z'$ have been determined at the 1$\sigma$ CL in
the $e^+e^-\to\mu^+\mu^-$ process, and at the 2$\sigma$ CL in the
Bhabha process. The $Z'$ mass was estimated to be 1--1.2 TeV. An
increase in statistics could make these signals more pronounced.
In Ref. \cite{Gulov:2007zz} the updated results of the
one-parameter fit and  the complete many-parametric fit of the
LEP2 data were performed  with the goal  to estimate a possible
signal of the $Z'$-boson with accounting for the final data of the
LEP collaborations DELPHI and OPAL
\cite{Abbiendi:2003dh,Abbiendi:1998ea,Ackerstaff:1997nf,Abdallah:2005ph}.
Usually, in a many-parametric fit the uncertainty of the result
increases drastically because of extra parameters. On the
contrary, in our approach due to the relations between $Z'$
couplings there are only 2-3 independent parameters for the
investigated leptonic  scattering processes. As it was  showed in
Ref. \cite{Gulov:2007zz}, an inevitable increase of confidence
areas in the many-parametric space was compensated due to
accounting for all accessible experimental information. Therefore,
the uncertainty of the many-parametric fit was estimated as the
comparable with previous one-parametric fits in Refs.
\cite{Demchik:2003pd,Gulov:2004sg}. In this approach the combined
data fit for all lepton processes is also possible. Note that the
hints for the $Z'$ have been determined in all the processes
considered that increases the reliability of the signal. These
results may serve as a good input into the LHC and  future ILC
experiments and used in various aspects. To underline the
importance of them we mention that there are many tools at the LHC
for the identification  of $Z'$. But many of them are only
applicable if $Z'$ is relatively light. The knowledge of the $Z'$
couplings to SM fermions also have important consequences. As
concerns the notion ``model-independent search'' used below, it
refers to a  class of models containing $Z'$  and inspired by the
grand-unified field theories. It does not mean all possible ones.
But if one determines the signal of this state, further
specification of the underlying theory could follow.

The paper is organized as follows. In sect. 2 we give a necessary
information about the description of $Z'$ at low energies and
introduce the relations between the $Z'$ couplings. In sect. 3 the
cross sections and the observables to pick out uniquely the
virtual $Z'$ in the $e^+ e^- \to \mu^+ \mu^-, \tau^+ \tau^-$
processes  are given. The fits of data are described and
discussed. Then in sect. 4 the same is present for the Bhabha
process $e^+ e^- \to e^+ e^-$. The one parametric and two
parametric fits are discussed.  In sect. 5 we discuss the role of
the present model-independent analysis for the LHC experiments.
The discussion and comparison with results of other approaches are
given in sect. 6.

\section{The Abelian $Z'$ boson at low energies}

Let us adduce a necessary information about the Abelian
$Z'$-boson. This particle is predicted by a number of grand
unification models. Among them the $E_6$ and $SO(10)$ based models
\cite{Hewett:1988xc} (for instance, LR, $\chi-\psi$ and so on) are
often discussed in the literature. In all the models, the Abelian
$Z'$-boson is described by a low-energy $\tilde{U}(1)$ gauge
subgroup originated in some symmetry breaking pattern.

At low energies, the $Z'$-boson can manifest itself by means of
the couplings to the SM fermions and scalars as a virtual
intermediate state. Moreover, the $Z$-boson couplings are also
modified due to a $Z$--$Z'$ mixing. In principle, arbitrary
effective $Z'$ interactions to the SM fields could be considered
at low energies. However, the couplings of non-renormalizable
types have to be suppressed by heavy mass scales because of
decoupling. Therefore, significant signals beyond the SM can be
inspired by the couplings of renormalizable types. Such couplings
can be derived by adding new $\tilde{U}(1)$-terms to the
electroweak covariant derivatives $D^\mathrm{ew}$ in the
Lagrangian \cite{Cvetic:1986ff,Degrassi:1989mu} (review, Ref.
\cite{Leike:1998wr,Langacker:2008yv})
\begin{equation}
\label{Lscal} L_\phi = \left| \left(
\partial_\mu - \frac{i g}{2} \sigma_a W^a_\mu - \frac{i g'}{2}
B_\mu Y_{\phi}  - \frac{i \tilde{g}}{2} \tilde{B}_\mu
\tilde{Y}_{\phi} \right) \phi\right|^2,
\end{equation}
\begin{eqnarray}
\label{Lf} L_f &=& i
\sum\limits_{f_L} \bar{f}_L \gamma^\mu \left(
\partial_\mu - \frac{i g}{2} \sigma_a W^a_\mu - \frac{i g'}{2}
B_\mu Y_{f_L} - \frac{i \tilde{g}}{2}\tilde{B}_\mu
\tilde{Y}_{f_L}\right) f_L  \nonumber\\ &+& i \sum\limits_{f_R}
\bar{f}_R \gamma^\mu \left(
\partial_\mu  - i
g' B_\mu Q_{f} - \frac{i \tilde{g}}{2}\tilde{B}_\mu
\tilde{Y}_{f_R}\right) f_R,
\end{eqnarray}
where summation over all the SM left-handed fermion doublets,
leptons and quarks, $f_L = {(f_u)_L, (f_d)_L}$, and the
right-handed singlets, $f_R = (f_u)_R, (f_d)_R $, is understood.
In these formulas, $g, g', \tilde{g}$ are the charges associated
with the $SU(2)_L, U(1)_Y,$ and  the $Z'$ gauge groups,
respectively, $\sigma_a$ are the Pauli matrices, $Q_f$ denotes the
charge of $f$ in positron charge units, $Y_{\phi}$ is the $U(1)_Y$
hypercharge, and $Y_{f_L}= -1$ for leptons and 1/3 for quarks. In
general, generators
$\tilde{Y}_{f_L}=\mathrm{diag}(\tilde{Y}_{f_u}, \tilde{Y}_{f_d})$
and $\tilde{Y}_{\phi} = \mathrm{diag}(\tilde{Y}_{\phi,{1}},
\tilde{Y}_{\phi,{2}})$ are diagonal $2\times 2$ matrices. As for
the scalar sector, the Lagrangian can be simply generalized for
the case of the SM with two light Higgs doublets (THDM).

The Lagrangian (\ref{Lscal}) leads to the $Z$--$Z'$ mixing. The
$Z$--$Z'$ mixing angle $\theta_0$ is determined by the coupling
$\tilde{Y}_\phi$ as follows
\begin{equation}\label{2}
\theta_0 =
\frac{\tilde{g}\sin\theta_W\cos\theta_W}{\sqrt{4\pi\alpha_\mathrm{em}}}
\frac{m^2_Z}{m^2_{Z'}} \tilde{Y}_\phi
+O\left(\frac{m^4_Z}{m^4_{Z'}}\right),
\end{equation}
where $\theta_W$ is the SM Weinberg angle, and
$\alpha_\mathrm{em}$ is the electromagnetic fine structure
constant. Although the mixing angle is a small quantity of order
$m^{-2}_{Z'}$, it contributes to the $Z$-boson exchange amplitude
and cannot be neglected at the LEP energies.

In what follows we will also use the $Z'$ couplings to the vector
and axial-vector fermion currents defined as
 \begin{equation} \label{av} v_f =
\tilde{g}\frac{\tilde{Y}_{L,f} + \tilde{Y}_{R,f}}{2}, \qquad a_f =
\tilde{g}\frac{\tilde{Y}_{R,f} - \tilde{Y}_{L,f}}{2}.
\end{equation}
The Lagrangian (\ref{Lf}) leads to the following interactions
between the fermions and the $Z$ and $Z'$ mass eigenstates:
\begin{eqnarray}\label{ZZplagr}
{\cal L}_{Z\bar{f}f}&=&\frac{1}{2} Z_\mu\bar{f}\gamma^\mu\left[
(v^\mathrm{SM}_{fZ}+\gamma^5 a^\mathrm{SM}_{fZ})\cos\theta_0
+(v_f+\gamma^5 a_f)\sin\theta_0 \right]f, \nonumber\\
{\cal L}_{Z'\bar{f}f}&=&\frac{1}{2} Z'_\mu\bar{f}\gamma^\mu\left[
(v_f+\gamma^5 a_f)\cos\theta_0
-(v^\mathrm{SM}_{fZ}+\gamma^5
a^\mathrm{SM}_{fZ})\sin\theta_0\right]f,
\end{eqnarray}
where $f$ is an arbitrary SM fermion state; $v^\mathrm{SM}_{fZ}$,
$a^\mathrm{SM}_{fZ}$ are the SM couplings of the $Z$-boson.

At low energies the $Z'$ couplings enter the cross-section
together with the inverse $Z'$ mass, so it is convenient to
introduce the dimensionless couplings
\begin{equation}\label{6}
\bar{a}_f=\frac{m_Z}{\sqrt{4\pi}m_{Z'}}a_f,\quad
\bar{v}_f=\frac{m_Z}{\sqrt{4\pi}m_{Z'}}v_f,
\end{equation}
which can be constrained by experiments.

The low energy parameters $\tilde{Y}_{\phi}$, $\tilde{Y}_{L,f}$,
$\tilde{Y}_{R,f}$ must be fitted in experiments. In most
investigations they were considered as independent ones. In a
particular model, the couplings $ \tilde{Y}_{\phi}$,
$\tilde{Y}_{L,f}$, $\tilde{Y}_{R,f}$ take some specific values. In
case when the model is unknown, these parameters  remain
potentially arbitrary numbers. However, this is not the case if
one assumes that the underlying extended model is a renormalizable
one.

In Refs. \cite{Gulov:2000eh,Gulov:1999ry} it was shown that these
parameters are correlated. This correlations follows if the
underlying unknown theory is a renormalizable one. The following
conditions were assumed to derive the relations between $Z'$
couplings:
\begin{enumerate}
\item
only one neutral vector boson exists at energy scale about 1-10
TeV,
\item
the $Z'$ boson can be phenomenologically described by the
effective Lagrangian (\ref{Lf}), (\ref{Lscal}) at low energies,
\item
the $Z'$ boson and other possible heavy particles are decoupled at
considered energies, and the theory beyond the $Z'$ decoupling
scale is either one- or two-Higgs-doublet standard model,
\item
the SM gauge group is a subgroup of possible extended gauge group
of the underlying theory. So, the only origin of possible
tree-level $Z'$ interactions to the SM vector bosons is the
$Z$--$Z'$ mixing.
\end{enumerate}
Under these conditions, we have obtained the relations between
phenomenological parameters of the effective Lagrangian
(\ref{Lf}), (\ref{Lscal}):
\begin{equation} \label{rgr2}
\tilde{Y}_{\phi,1} = \tilde{Y}_{\phi,2} \equiv \tilde{Y}_{\phi},
\qquad \tilde{Y}_{L,f}= \tilde{Y}_{L,f^*},\qquad \tilde{Y}_{R,f} =
\tilde{Y}_{L,f} + 2 T_{3f}~ \tilde{Y}_{\phi}.
\end{equation}
Here $f$ and $f^*$ are the partners of the $SU(2)_L$ fermion
doublet ($l^* = \nu_l, \nu^* = l, q^*_u = q_d$ and $q^*_d = q_u$),
$T_{3f}$ is the third component of weak isospin. They are key point for
investigations reported below.

Introducing the $Z'$ couplings to the vector and axial-vector
fermion currents (\ref{av}), the last formula in Eq. (\ref{rgr2})
yields
\begin{equation} \label{grgav}
v_f - a_f= v_{f^*} - a_{f^*}, \qquad a_f = T_{3f}
\tilde{g}\tilde{Y}_\phi.
\end{equation}
The couplings of the Abelian $Z'$ to the axial-vector fermion
current have a universal absolute value proportional to the $Z'$
coupling to the scalar doublet. Then, the $Z$--$Z'$ mixing angle
(\ref{2}) can be determined by the axial-vector coupling. As a
result, the number of independent couplings is significantly
reduced.

We assume no new light particles. The relations could change
essentially if the SM has to be modified at energies below the
$Z'$ mass.

The derived relations are necessary but not exhaustive constraints
on the $Z'$ couplings. To derive
exhaustive constraints one need to fix the complete particle
content at high energies in order to ensure the cancelation of
ultraviolet divergencies in arbitrary scattering process.

The relations (\ref{grgav}) were derived for effective low-energy
parameters accounting for radiation corections. Nevertheless, they
also hold at tree-level in a wide class of known models containing
the Abelian $Z'$. In this case, it is  possible to derive them  by
imposing the requirement that the SM Lagrangian (including Yukawa
term) has to be invariant with respect to the extra $\tilde{U}(1)$
group associated with the $Z'$ \cite{Gulov:2001ia}.

A lot widely  discussed models are derived from
the ${\rm E}_6$ group (the so called LR, $\chi$-$\psi$ models).
The tree-level $Z'$ couplings to the SM fermions in the models are
shown in Table \ref{zpcoup}.

\begin{table}[ph]
\caption{The $Z^\prime$ couplings to the SM fermions in the most
discussed ${\rm E}_6$-based models.}
{\begin{tabular}{@{}c|cc|cc@{}} \hline $f$ &
\multicolumn{2}{|c|}{$\chi$-$\psi$} &
\multicolumn{2}{|c}{LR}\\
 \cline{2-5}
 & $a_f/\tilde{g}$ & $v_f/\tilde{g}$
      & $a_f/\tilde{g}$ & $v_f/\tilde{g}$
\\ \hline
\rule{0pt}{15pt} $\nu$ &
 $-3\frac{\cos{\beta}}{\sqrt{40}}
  -\frac{\sin{\beta}}{\sqrt{24}}$ &
 $3\frac{\cos{\beta}}{\sqrt{40}}
  +\frac{\sin{\beta}}{\sqrt{24}}$ &
 $-\frac{1}{2\alpha}$ & $\frac{1}{2\alpha}$ \\
\rule{0pt}{15pt}  $e$  &
 $-\frac{\cos{\beta}}{\sqrt{10}}
  -\frac{\sin{\beta}}{\sqrt{6}}$ &
 $2\frac{\cos{\beta}}{\sqrt{10}}$ &
 $-\frac{\alpha}{2}$ & $\frac{1}{\alpha}-\frac{\alpha}{2}$ \\
\rule{0pt}{15pt}  $q_u$ &
 $\frac{\cos{\beta}}{\sqrt{10}}
 -\frac{\sin{\beta}}{\sqrt{6}}$ & 0 &
 $\frac{\alpha}{2}$&$-\frac{1}{3\alpha}+\frac{\alpha}{2}$ \\
\rule{0pt}{15pt}  $q_d$ &
 $-\frac{\cos{\beta}}{\sqrt{10}}
  -\frac{\sin{\beta}}{\sqrt{6}}$ &
 $-2\frac{\cos{\beta}}{\sqrt{10}}$ &
 $-\frac{\alpha}{2}$ & $-\frac{1}{3\alpha}-\frac{\alpha}{2}$
\\
 \hline
\end{tabular}\label{zpcoup}}
\end{table}

The ${\rm E}_6$-symmetry breaking scheme
\[
{\rm E}_6\to{\rm SO}(10)\times{\rm U}(1)_\psi,\quad {\rm
SO}(10)\to{\rm SU}(3)_c\times{\rm SU}(2)_L \times{\rm
SU}(2)_R\times{\rm U}(1)_{B-L}.
\]
leads to the so called left-right (LR) model. Another scheme,
\[
{\rm E}_6\to{\rm SO}(10)\times{\rm U}(1)_\psi\to{\rm SU}(5)\times
{\rm U}(1)_\chi\times{\rm U}(1)_\psi,
\]
predicts the Abelian $Z^\prime$ which is a linear combination of
the neutral vector bosons $\psi$ and $\chi$,
\[
Z^\prime=\chi\cos{\beta} +\psi\sin{\beta}
\]
with the mixing angle $\beta$. If we suppose only one $Z'$ boson
at low energies, the $\psi$ boson should be much heavier than the
$\chi$ field. In this case the field $\psi$ is decoupled and
$\beta\to 0$. As it is seen, both the LR and the $\chi$-$\psi$
models (with $\beta=0$ to avoid two $Z'$ bosons with the same
scale of masses) satisfy the relations (\ref{rgr2}) except for
neutrinos. This fact is a consequence of the zero neutrino mass
assumed already.
 Neutrinos are not detected
in the discussed experiments, so the question about $Z'$
interactions to neutrinos is  inessential.

The relations (\ref{rgr2}) are valid not only for the $E_6$ based
models in Table \ref{zpcoup}. For example, the relations also
cover the Sequential SM (SSM) mentioned in reports of LEP
Collaborations. Thus, they describe correlations between $Z'$
couplings for a wide set of models beyond the SM. That is the
reason to call the relations model-independent ones.

LEP collaborations have applied model dependent search for $Z'$
and obtained the low bounds on the mass $m_{Z'} \geq 400 - 800$
GeV dependently on a specific model
\cite{Alcaraz:2006mx,Abbiendi:2003dh,Abbiendi:1998ea,Ackerstaff:1997nf,Abdallah:2005ph}.
In our analysis, the relations (\ref{rgr2}) give a possibility to
reduce the number of fitted parameters, to determine kinematics of
the processes, and to introduce observables which uniquely pick
out the $Z'$ signals. Therefore we are able to distinguish  the
particle instead of constraining its mass.

\section{$Z'$ search in $e^+e^-\to\mu^+\mu^-,\tau^+\tau^-$ processes}

\subsection{The differential cross section}

Let us  consider the processes $e^+e^-\to l^+l^-$
($l=\mu,\tau$) with the non-polarized initial and final state
fermions. In order to introduce the observable which selects the
signal of the Abelian $Z'$ boson we need to compute the
differential cross-sections of the processes up to the one-loop
level. Two classes of Feynman diagrams are taken into account. The
first one includes the pure SM graphs. The set of SM diagrams give
the SM prediction for the process which is a background for
observation of possible deviations due to $Z'$ boson. Obviously,
the SM has to be estimated as accurate as possible. So, the full
set of radiative corrections must be taken into account. They are
the mass operators, the vertex corrections, the boxes, and the
effects of initial and final state radiation of soft photons. The
kinematic region allowed by the detectors is also important. Fine
cancelations of ultraviolet and infrared divergencies occur due to
renormalizability of the SM giving the finite result. The
resulting SM value is published by the Collaboration and have been
checked.

The second class of diagrams includes heavy $Z'$ boson as a
virtual state. Such graphs lead to small corrections of order
$s/m^2_{Z'}$ to cross section. Since the effective low energy
Lagrangian is used to describe $Z'$ interactions to SM particles
and the particle content of the underlying theory remains hidden,
one has to consider the $Z'$ contribution in the decoupling limit.
Namely, we assume that $Z'$ is not excited inside loops. The
tree-level diagram $e^+e^-\to Z'\to l^+l^-$ defines a leading
contribution to the cross-section. It is enough to take into
account this diagram to estimate possible $Z'$ signals. The
cross-section includes the interference of the $Z'$-exchange
amplitude with the SM amplitudes. Radiative corrections were
incorporated with the $Z'$-exchange diagram in the improved Born
approximation.

In actual calculations and experimental data treating, the SM
values of cross sections coincide with the results of the LEP
Collaborations and the deviations due to the $Z'$ boson have been
computed in the improved Born approximation at one-loop level. The
same approach is used for the Bhabha process which will be
analyzed in next section. This is sufficient to analyze the
present day experimental data.

In the lower order in $m^{-2}_{Z'}$ the $Z'$ contributions to the
differential cross-section of the process $e^+e^-\to l^+l^-$ are
expressed in terms of four-fermion contact couplings, only. If one
takes into consideration the higher-order corrections in
$m^{-2}_{Z'}$, it becomes possible to estimate separately the
$Z'$-induced contact couplings and the $Z'$ mass
\cite{Rizzo:1996rx}. In the present analysis we keep the terms of
order $O(m^{-4}_{Z'})$ to fit both of these parameters.

Expanding the differential cross-section in the inverse $Z'$ mass
and neglecting the terms of order $O(m^{-6}_{Z'})$, we have
\begin{eqnarray}
\frac{d\sigma_l(s)}{dz} &=& \frac{d\sigma_l^{\rm SM}(s)}{dz}
+\sum_{i=1}^{7}\sum_{j=1}^{i}
\left[A_{ij}^l(s,z)+B_{ij}^l(s,z)\zeta\right]x_{i}x_{j}
\nonumber\\&&
+\sum_{i=1}^{7}\sum_{j=1}^{i}\sum_{k=1}^{j}\sum_{n=1}^{k}
C_{ijkn}^l(s,z)x_{i}x_{j}x_{k}x_{n},
\end{eqnarray}
where the dimensionless quantities
\begin{eqnarray}
 \zeta = \frac{m^2_Z}{m^2_{Z'}},\quad
 (x_1,x_2,x_3,x_4,x_5,x_6,x_7) =
 (\bar{a},\bar{v}_e,\bar{v}_\mu,\bar{v}_\tau,\bar{v}_d,\bar{v}_s,\bar{v}_b)
\end{eqnarray}
are introduced. Since the axial-vector couplings of the Abelian
$Z'$ boson are universal, we use the shorthand notation
$\bar{a}=\bar{a}_e$. In what follows the index $l=\mu,\tau$
denotes the final-state lepton.

The coefficients $A$, $B$, $C$ are determined by the SM couplings
and masses. Each factor may include the tree-level contribution,
the one-loop correction and the term describing the soft-photon
emission. The factors $A$ describe the leading-order contribution,
whereas others correspond to the higher order corrections in
$m^{-2}_{Z'}$.

\subsection{The observable}

To take into consideration the correlations (\ref{2}) we introduce
the observable $\sigma_l(z)$ defined as the difference of cross
sections integrated in some ranges of the scattering angle
$\theta$ \cite{Gulov:1999ry,Demchik:2003pd}:
\begin{eqnarray}\label{eq8}
 \sigma_l(z)
 &\equiv&\int\nolimits_z^1
  \frac{d\sigma_l}{d\cos\theta}d\cos\theta
 -\int\nolimits_{-1}^z
  \frac{d\sigma_l}{d\cos\theta}d\cos\theta,
\end{eqnarray}
where $z$ stands for the cosine of the boundary angle. The idea of
introducing the $z$-dependent observable (\ref{eq8}) is to choose
the value of the kinematic parameter $z$ in such a way that to
pick up the characteristic features of the Abelian $Z'$ signals.

The deviation of the observable from its SM value can be derived
by the angular integration of the differential cross-section and
has the form:
\begin{eqnarray}
\Delta\sigma_l(z) &=& \sigma_l(z) - \sigma^{\rm SM}_l(z)
=\sum_{i=1}^{7}\sum_{j=1}^{i}
\left[\tilde{A}_{ij}^l(s,z)+\tilde{B}_{ij}^l(s,z)\zeta\right]x_{i}x_{j}
\nonumber\\&&
+\sum_{i=1}^{7}\sum_{j=1}^{i}\sum_{k=1}^{j}\sum_{n=1}^{k}
\tilde{C}_{ijkn}^l(s,z)x_{i}x_{j}x_{k}x_{n}.
\end{eqnarray}

\begin{figure}
\centerline{\psfig{file=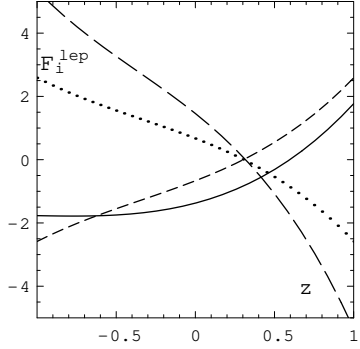,width=4.7cm}} \vspace*{8pt}
 \caption{The leptonic functions
  $F^{l}_{0}$ (the solid curve),
  $F^{l}_{1}$ (the long-dashed curve),
  $F^{l}_{2}$ (the dashed curve), and
  $F^{l}_{3}$ (the dotted curve) at $\sqrt{s}=500$ GeV.}
 \label{fig:2f}
\end{figure}

Then let us introduce the quantity
$\Delta\sigma\left(z\right)\equiv \sigma\left(z\right)
-{\sigma}_{SM}\left(z\right)$ which owing to the relations
(\ref{rgr2}) can be written in the form
\begin{eqnarray}\label{obs:6}
 \Delta\sigma_f(z)
 &=&\frac{\alpha N_f}{8}\frac{g^2_{Z^\prime}}{m^2_{Z^\prime}}
  \left[
  F^f_0(z,s) \tilde{Y}^2_\phi
  +2 F^f_1(z,s) T_{3f}\tilde{Y}_{L,f}\tilde{Y}_{L,e}  \right.
 \nonumber\\
  &&
  +\left.
  2 F^f_2(z,s) T_{3f}\tilde{Y}_{L,f}\tilde{Y}_\phi
  + F^f_3(z,s) \tilde{Y}_{L,e}\tilde{Y}_\phi \right].
\end{eqnarray}
The factor functions $F^f_i(z,s)$ depend on the fermion type
through the $|Q_f|$, only. In Fig. \ref{fig:2f} they are shown as
the functions of $z$ for $\sqrt{s}=500$ GeV. The leading
contributions to $F^f_i(z,s)$,
\begin{eqnarray}\label{obs:7}
 F^f_0(z,s)&=&
  -\frac{4}{3}\left|Q_f\right|
  \left(1 -z -z^{2} -\frac{z^{3}}{3}\right)
+O\left(\frac{m^2_Z}{s}\right),
 \nonumber\\
 F^f_1(z,s)&=&\frac{4}{3}
   \left[1 -z^{2} -\left|Q_f\right|
   \left(3z+z^{3}\right)\right]
+O\left(\frac{m^2_Z}{s}\right),
 \nonumber\\
 F^f_2(z,s)&=&
  -\frac{2}{3}\left(1-z^{2}\right)
   +\frac{2}{9}\left(3z+z^{3}\right)
   \left(4\left|Q_f\right|-1\right)
      +O\left(\frac{m^2_Z}{s}\right),
 \nonumber\\
 F^f_3(z,s)&=&\frac{2}{3}\left|Q_f\right|
   \left(1-3z-z^{2}-z^{3}\right)
+O\left(\frac{m^2_Z}{s}\right),
\end{eqnarray}
are given by the $Z^\prime$ exchange diagram $e^-e^+\to
Z'\to\bar{f}f$, since the $Z$--$Z'$ mixing contribution to the $Z$
exchange diagram is suppressed by the factor $m^2_Z/s$.

From Eqs. (\ref{obs:7}) one can see that the leading contributions
to the leptonic factors $F^l_1$, $F^l_2$, $F^l_3$ are found to be
proportional to the same polynomial in $z$. This is the
characteristic feature of the leptonic functions $F^l_i$
originating due to the kinematic properties of fermionic currents
and the specific values of the SM leptonic charges. Therefore, it
is possible to choose the value of $z=z^\ast$ which switches off
three leptonic factors $F^l_1$, $F^l_2$, $F^l_3$ simultaneously.
Moreover, the quark function $F^q_3$ in the lower order is
proportional to the leptonic one and therefore is switched off,
too. As is seen from Fig. \ref{fig:2f}, the appropriate value of
$z^\ast$ is about $\sim 0.3$. By choosing this value of $z^\ast$
one can simplify Eq. (\ref{obs:6}). It is also follows from Eq.
(\ref{obs:6}) that neglecting the factors $F^l_1$, $F^l_2$,
$F^l_3$ one obtains the sign definite quantity $\Delta \sigma_l
(z^\ast)\sim \tilde{Y}^2_\phi\sim\bar{a}^2$.

There is the interval of   boundary angle values at which the
factors $\tilde{A}^l_{11}$, $\tilde{B}^l_{11}$, and
$\tilde{C}^l_{1111}$ at the sign-definite parameters $\bar{a}^2$,
$\bar{a}^2\zeta$, and $\bar{a}^4$ contribute more than 95\% of the
observable value. It gives a possibility to construct the
sign-definite observable $\Delta\sigma_l(z^*)<0$ by specifying the
proper value of $z^*$.

In general, one could choose the boundary angle $z^*$ in different
schemes. If just a few number of tree-level four-fermion contact
couplings are considered, one can specify $z^*$ in order to cancel
the factor at the vector-vector coupling. However, if one-loop
corrections are taken into account there is a large amount of
additional contact couplings. So, we have to define some
quantitative criterion $F(z)$ to estimate the contributions from
sign-definite factors at a given value of the boundary angle $z$.
Maximizing the criterion, one could derive the value $z^*$
corresponding to the sign-definite observable
$\Delta\sigma_l(z^*)$. Since the observable is linear in the
coefficients $A$, $B$, and $C$, we introduce the following
criterion,
\begin{equation}
F=\frac{|\tilde{A}_{11}|+\omega_B |\tilde{B}_{11}| + \omega_C
|\tilde{C}_{1111}|}
 {\sum\limits_{{\rm all}~\tilde{A}}\left|\tilde{A}_{ij}\right|
  +\omega_B\sum\limits_{{\rm all}~\tilde{B}}\left|\tilde{B}_{ij}\right|
  +\omega_C\sum\limits_{{\rm all}~\tilde{C}}\left|\tilde{C}_{ijkn}\right|},
\end{equation}
where the positive `weights' $\omega_B\sim\zeta$ and
$\omega_C\sim\epsilon$ take into account the order of each term in
the inverse $Z'$ mass.

The numeric values of the `weights' $\omega_B$ and $\omega_C$ can
be taken from the present day bounds on the contact couplings
\cite{Alcaraz:2006mx}. As the computation shows, the value of
$z^*$ with the accuracy $10^{-3}$ depends on the order of the
`weight' magnitudes, only. So, in what follows we take
$\omega_B\sim 4\times 10^{-3}$ and $\omega_C\sim 4\times 10^{-5}$.

The function $z^\ast(s)$ is the decreasing function of the
center-of-mass energy. It is tabulated for the LEP2 energies in
Table \ref{tobsmu}. The corresponding values of the maximized
function $F$ are within the interval $0.95<F<0.96$.

\begin{table}[ph]
\caption{The boundary angle $z^*$ and the coefficients in the
observable $\Delta\sigma_l(z^*)$ for the scattering into $\mu$ and
$\tau$ pairs at the one-loop level.}
{\begin{tabular}{@{}c|cc|cc@{}}\hline
 $\sqrt{s}$, GeV & \multicolumn{2}{|c|}{$z^*$}
 & \multicolumn{2}{|c}{$\tilde{A}_{11}$,$\times 10^{2}$}
\\ \cline{2-5}
  & $\mu^+\mu^-$ & $\tau^+\tau^-$ & $\mu^+\mu^-$ & $\tau^+\tau^-$
\\ \hline
 130 & 0.450 & 0.460 & -7.29 & -6.87 \\
 136 & 0.439 & 0.442 & -7.09 & -6.88 \\
 161 & 0.400 & 0.400 & -6.43 & -6.25 \\
 172 & 0.390 & 0.391 & -6.19 & -6.01 \\
 183 & 0.383 & 0.385 & -5.99 & -5.71 \\
 189 & 0.380 & 0.380 & -5.86 & -5.68 \\
 192 & 0.380 & 0.380 & -5.79 & -5.62 \\
 196 & 0.380 & 0.379 & -5.71 & -5.54 \\
 200 & 0.378 & 0.378 & -5.64 & -5.47 \\
 202 & 0.376 & 0.377 & -5.60 & -5.43 \\
 205 & 0.374 & 0.374 & -5.55 & -5.48 \\
 207 & 0.372 & 0.372 & -5.52 & -5.44 \\
 \hline
\end{tabular} \label{tobsmu}}
\end{table}

Since $\tilde{A}^l_{11}(s,z^*)<0$, $\tilde{B}^l_{11}(s,z^*)<0$ and
$\tilde{C}^l_{1111}(s,z^*)<0$, the observable
\begin{equation}\label{7dfg}
 \Delta\sigma_l(z^*)=
 \left[\tilde{A}^l_{11}(s,z^*) +\zeta\tilde{B}^l_{11}(s,z^*)
 \right]\bar{a}^2 + \tilde{C}^l_{1111}(s,z^*)\bar{a}^4
\end{equation}
is negative with the accuracy 4--5\%. Since this property follows
from the relations (\ref{grgav}) for the Abelian $Z'$ boson, the
observable $\Delta\sigma_l(z^*)$ selects the model-indepen\-dent
signal of this particle in the processes $e^+e^-\to l^+l^-$. It
allows to use the data on scattering into $\mu\mu$ and $\tau\tau$
pairs in order to estimate the Abelian $Z'$ coupling to the
axial-vector lepton currents.

Although the observable can be computed from the differential
cross-sections directly, it is also possible to recalculate it
from the total cross-sections and the forward-backward
asymmetries. The recalculation procedure has the proper
theoretical accuracy. Nevertheless, it allows to reduce the
experimental errors on the observable, since the published data on
the total cross-sections and the forward-backward asymmetries are
more precise than the data on the differential cross-sections.

The recalculation is based on the fact that the differential
cross-section can be approximated with a good accuracy by the
two-parametric polynomial in the cosine of the scattering angle
$z$:
\begin{equation}
\frac{d\sigma_l(s)}{dz} = \frac{d\sigma_l^{\rm SM}(s)}{dz} +
(1+z^2)\beta_l + z \eta_l +\delta_l(z),
\end{equation}
where $\delta_l(z)$ measures the difference between the exact and
the approximated cross-sections. The approximated cross-section
reproduces the exact one in the limit of the massless initial- and
final-state leptons and if one neglects the contributions of the
box diagrams. Detailed analysis of this point is given in
\cite{Demchik:2003pd} where it was showed that the the theoretical
error is one order less than the corresponding statistical
uncertainty for the observable. Thus, the proposed approximation
is quite good and can be successfully used to obtain more accurate
experimental values of the observable.

\subsection{Data fit}

To search for the model-independent signals of the Abelian
$Z'$-boson we will analyze the introduced observable
$\Delta\sigma_l (z^\ast)$  (\ref{7dfg}) on the base of the LEP2 data set. In the
lower order in $m^{-2}_{Z'}$ it  depends
on one flavor-independent parameter $\bar{a}^2$,
\begin{equation}
 \Delta\sigma^{\rm th}_l(z^*)=
 \tilde{A}^l_{11}(s,z^*)\bar{a}^2 + \tilde{C}^l_{1111}(s,z^*)\bar{a}^4,
\end{equation}
which can be fitted from the experimental values of
$\Delta\sigma_\mu (z^\ast)$ and $\Delta\sigma_\tau (z^\ast)$. As
we axplained above, the sign of the fitted parameter ($\bar{a}^2 >0$)
is the characteristic feature of the Abelian $Z'$ signal.

In what follows we will apply the usual fit method based on the
likelihood function. The central value of $\bar{a}^2$ is obtained
by the minimization of the $\chi^2$-function:
\begin{equation}
\chi^2(\bar{a}^2) = \sum_{n} \frac{\left[\Delta\sigma^{\rm
ex}_{\mu,n}(z^*)- \Delta\sigma^{\rm th}_\mu(z^*)\right]^2}
{\delta\sigma^{\rm ex}_{\mu,n}(z^*)^2},
\end{equation}
where the sum runs over the experimental points entering a data
set chosen. The $1\sigma$ CL interval $(b_1,b_2)$ for the fitted
parameter is derived by means of the likelihood function ${\cal
L}(\bar{a}^2)\propto\exp[-\chi^2(\bar{a}^2)/2]$. It is determined
by the equations:
\begin{equation}
\int\nolimits_{b_1}^{b_2}{\cal L}(\epsilon ')d\epsilon ' = 0.68,
 \quad
{\cal L}(b_1)={\cal L}(b_2).
\end{equation}

To relate our results with those of Refs. \cite{Alcaraz:2006mx} we
introduce the contact interaction scale
\begin{equation}
\Lambda^2 = 4m^2_Z\bar{a}^{-2}.
\end{equation}
This normalization of contact couplings is admitted in Refs.
\cite{Alcaraz:2006mx}. We use again the likelihood method to
determine a one-sided lower limit on the scale $\Lambda$ at the
95\% CL. It is derived by the integration of the likelihood
function over the physically allowed region $\bar{a}^2>0$. The
strict definition is
\begin{equation}
\Lambda=2m_Z (\epsilon^*)^{-1/2}, \quad \int_{0}^{\epsilon^*}{\cal
L}(\epsilon ')d\epsilon ' = 0.95\int_{0}^{\infty}{\cal L}(\epsilon
')d\epsilon '.
\end{equation}

We also introduce the probability of the Abelian $Z'$ signal as
the integral of the likelihood function over the positive values
of $\bar{a}^2$:
\begin{equation}
P=\int\nolimits_{0}^{\infty} L(\epsilon ')d\epsilon '.
\end{equation}

Actually, the fitted value of the contact coupling $\bar{a}^2$
originates mainly from the leading-order term in the inverse $Z'$
mass in Eq. (\ref{7dfg}). The analysis of the higher-order terms
allows to estimate the constraints on the $Z'$ mass alone.
Substituting $\bar{a}^2$ in the observable (\ref{7dfg}) by its
fitted central value, one obtains the expression
\begin{equation}
 \Delta\sigma_l(z^*)=
 \left[\tilde{A}^l_{11}(s,z^*)
  +\zeta \tilde{B}^l_{11}(s,z^*) \right]\bar{a}_\mathrm{fitted}^2 +
 \tilde{C}^l_{1111}(s,z^*)\bar{a}_\mathrm{fitted}^4,
\end{equation}
which depends on the parameter $\zeta=m^2_Z/m^2_{Z'}$. Then, the
central value of this parameter and the corresponding 1$\sigma$ CL
interval are derived in the same way as those for $\bar{a}^2$.

To fit the parameters $\bar{a}^2$ and $\zeta$ we start with the
LEP2 data on the total cross-sections and the forward-backward
asymmetries \cite{Alcaraz:2006mx}. The corresponding values of the
observable $\Delta\sigma_l(z^\ast)$ with their uncertainties
$\delta\sigma_l(z^\ast)$ are calculated from the data by means of
the following relations:
\begin{eqnarray}
 \Delta\sigma_l(z^\ast)
 &=&
 \left[
 A_l^{\rm FB}\left(1-z^{\ast 2}\right)
 -\frac{z^\ast}{4}\left(3 +z^{\ast 2}\right)
 \right] \Delta\sigma_l^{\rm T}
 + \left(1 - z^{\ast 2}\right)
  \sigma_l^{\rm T,SM} \Delta A_l^{\rm FB},
 \\ \nonumber
 \delta\sigma_l(z^\ast)^2
 &=&
 {\left[
 A_l^{\rm FB}\left(1-z^{\ast 2}\right)
 -\frac{z^\ast}{4}\left(3 +z^{\ast 2}\right)
 \right]}^2 (\delta\sigma_l^{\rm T})^2
 +{\left[
 \left(1 - z^{\ast 2}\right)
 \sigma_{l}^{\rm T,SM}
 \right]}^2 (\delta A_l^{\rm FB})^2.
\end{eqnarray}

We perform the fits assuming several data sets, including the
$\mu\mu$, $\tau\tau$, and the complete $\mu\mu$ and $\tau\tau$
data, respectively. The results are presented in Table
\ref{tfitrecalc}.
\begin{table}
\caption{The contact coupling $\bar{a}^2$ with the 68\% CL
uncertainty, the 95\% CL lower limit on the scale $\Lambda$, the
probability of the $Z'$ signal, $P$, and the value of
$\zeta=m^2_Z/m^2_{Z'}$ as a result of the fit of the observable
recalculated from the total cross-sections and the
forward-backward asymmetries. }{\label{tfitrecalc}
\begin{tabular}{@{}lcccc@{}}
\hline Data set & $\bar{a}^2$, $\times 10^{-5}$ & $\Lambda$, TeV &
$P$ & $\zeta$, $\times 10^{-2}$
 \\ \hline
\rule{0pt}{12pt} $\mu\mu$ & $3.66^{+4.89}_{-4.86}$
 & 16.4
 & 0.77
 & $0.9\pm 27.8$
 \\
\rule{0pt}{12pt} $\tau\tau$ & $-2.66^{+6.43}_{-6.39}$
 & 17.4
 & 0.34
 & $-0.1\pm 50.1$
 \\
\rule{0pt}{12pt} $\mu\mu$ and $\tau\tau$ & $1.33^{+3.89}_{-3.87}$
 & 19.7
 & 0.63
 & $1.7\pm 60.9$
 \\ \hline
\end{tabular}}
\end{table}
As is seen, the more precise $\mu\mu$ data demonstrate the signal
of about 1$\sigma$ level. It corresponds to the Abelian $Z'$-boson
with the mass of order 1.2--1.5 TeV if one assumes the value of
$\tilde\alpha=\tilde{g}^2/4\pi$ to be in the interval 0.01--0.02.
No signal is found by the analysis of the $\tau\tau$
cross-sections. The combined fit of the $\mu\mu$ and $\tau\tau$
data leads to the signal below the 1$\sigma$ CL.

Being governed by the next-to-leading contributions in
$m^{-2}_{Z'}$, the fitted values of $\zeta$ are characterized by
significant errors. The $\mu\mu$ data set gives the central value
which corresponds to $m_{Z'}\simeq 1.1$ TeV.

We also perform a separate fit of the parameters based on the
direct calculation of the observable from the differential
cross-sections. The experimental uncertainties of the data on the
differential cross-sections are of one order larger than the
corresponding errors of the total cross-sections and the
forward-backward asymmetries. These data also provide the larger
values of the contact coupling $\bar{a}^2$. As for the more
precise $\mu\mu$ data, three of the LEP2 Collaborations
demonstrate positive values of $\bar{a}^2$. The combined
$\bar{a}^2$ is also positive and remains practically unchanged by
the incorporation of the $\tau\tau$ data.

\section{Search for $Z'$ in $e^+e^-\to e^+e^-$ process}

\subsection{The differential cross-section}

In our analysis of the Bhabha process, as the SM values of the cross-sections we use the
quantities calculated by the LEP2 collaborations
\cite{Abbiendi:2003dh,Abbiendi:1998ea,Ackerstaff:1997nf,Abdallah:2005ph,Barate:1999qx,Acciarri:1999rw}.
They account for either the one-loop radiative corrections or
initial and final state radiation effects (together with the event
selection rules, which are specific for each experiment). As it is
reported by the DELPHI Collaboration, there is a theoretical error
of the SM values of about 2\%. In our analysis this error is added
to the statistical and systematic ones for all the Collaborations.
As it was checked, the fit results are practically insensitive to
accounting for this error.

The deviation from the SM is computed in the improved Born
approximation. This approximation is sufficient for our analysis
leading to the systematic error of the fit results less than 5-10
per cents.

\begin{figure}
\centerline{\psfig{file=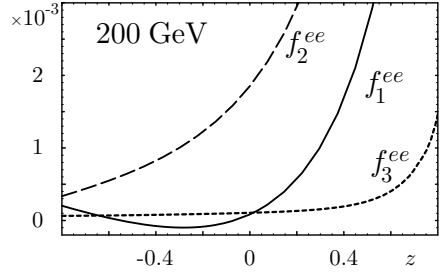,width=6cm}} \vspace*{8pt}
\caption{The factors at the $Z'$ couplings in the differential
cross-section of the Bhabha process.} \label{fig:0}
\end{figure}

The deviation from the SM of the differential cross-section for
the process $e^+e^-\to\ell^+\ell^-$ can be expressed through
various quadratic combinations of couplings $a=a_e$, $v_e$,
$v_\mu$, $v_\tau$. For the Bhabha process it reads
\begin{equation}\label{4}
\frac{d\sigma}{dz}-\frac{d\sigma^\mathrm{SM}}{dz} =
f^{ee}_1(z)\frac{a^2}{m_{Z'}^2} +
f^{ee}_2(z)\frac{v_e^2}{m_{Z'}^2} + f^{ee}_3(z)\frac{a
v_e}{m_{Z'}^2},
\end{equation}
where the factors are known functions of the center-of-mass energy
and the cosine of the electron scattering angle $z$ plotted in
Fig. \ref{fig:0}. The deviation of the cross-section for
$e^+e^-\to\mu^+\mu^-$ ($\tau^+\tau^-$) processes has a similar
form
\begin{eqnarray}\label{5}
\frac{d\sigma}{dz}-\frac{d\sigma^\mathrm{SM}}{dz} &=&
f^{\mu\mu}_1(z)\frac{a^2}{m_{Z'}^2} + f^{\mu\mu}_2(z)\frac{v_e
v_\mu}{m_{Z'}^2}
+ f^{\mu\mu}_3(z)\frac{a
v_e}{m_{Z'}^2} + f^{\mu\mu}_4(z)\frac{a v_\mu}{m_{Z'}^2}.
\end{eqnarray}
Eqs. (\ref{4})--(\ref{5}) are our definition of the $Z'$ signal.

Note again that the cross-sections in Eqs. (\ref{4})--(\ref{5})
account for the relations (\ref{rgr2}) through the functions
$f_1(z)$, $f_3(z)$, $f_4(z)$, since the coupling $\tilde{Y}_\phi$
(the mixing angle $\theta_0$) is substituted by the axial coupling
constant $a$. Usually, when a four-fermion effective Lagrangian is
applied to describe physics beyond the SM \cite{Babich:2002jb},
this dependence on the scalar field coupling is neglected at all.
However, in our analysis when we are interested in searching for
signals of the $Z'$-boson on the base of the effective low-energy
Lagrangian (\ref{Lf})--(\ref{Lscal}), these contributions to the
cross-section are essential.

\begin{figure}
\centerline{\psfig{file=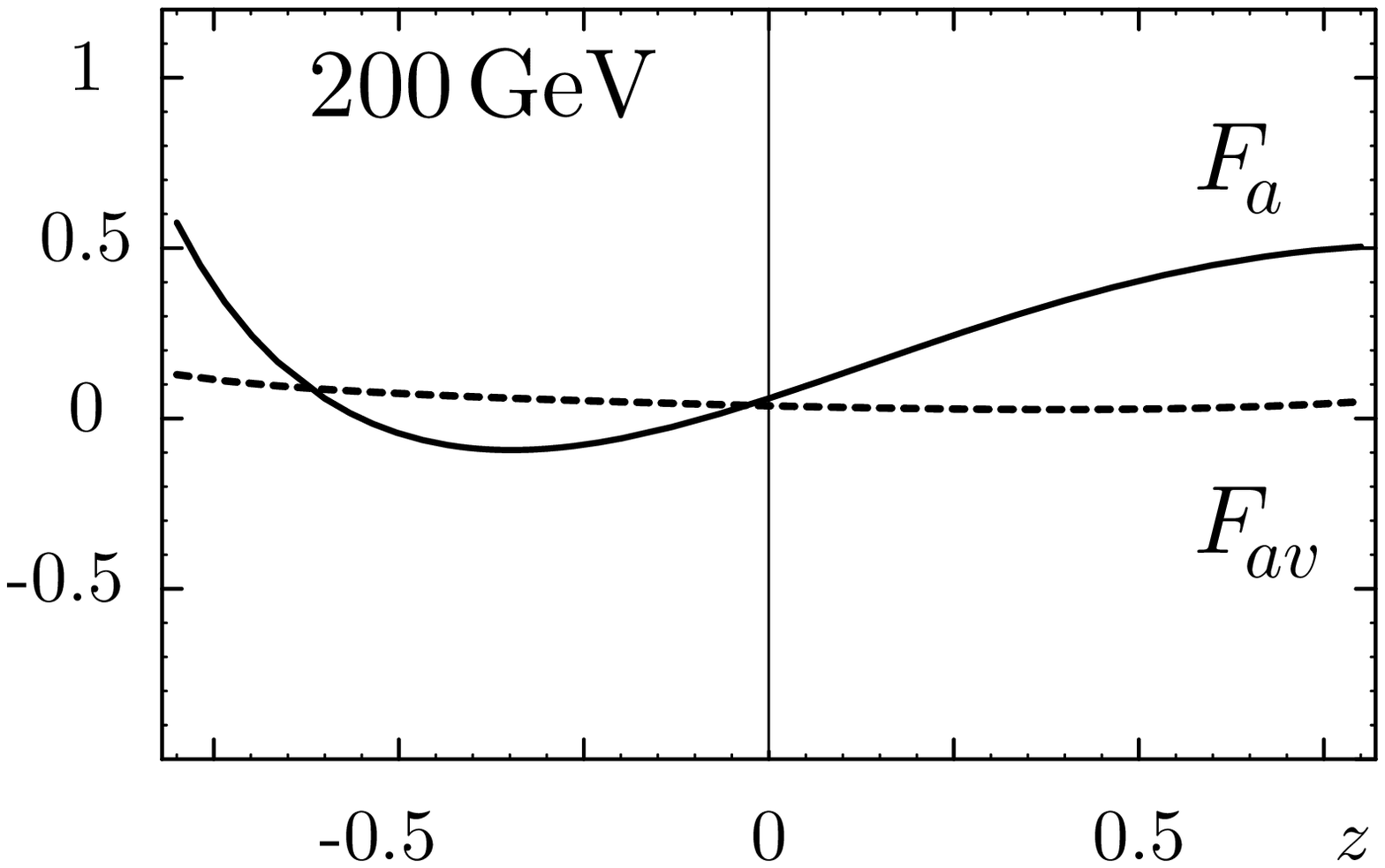,width=6cm}\psfig{file=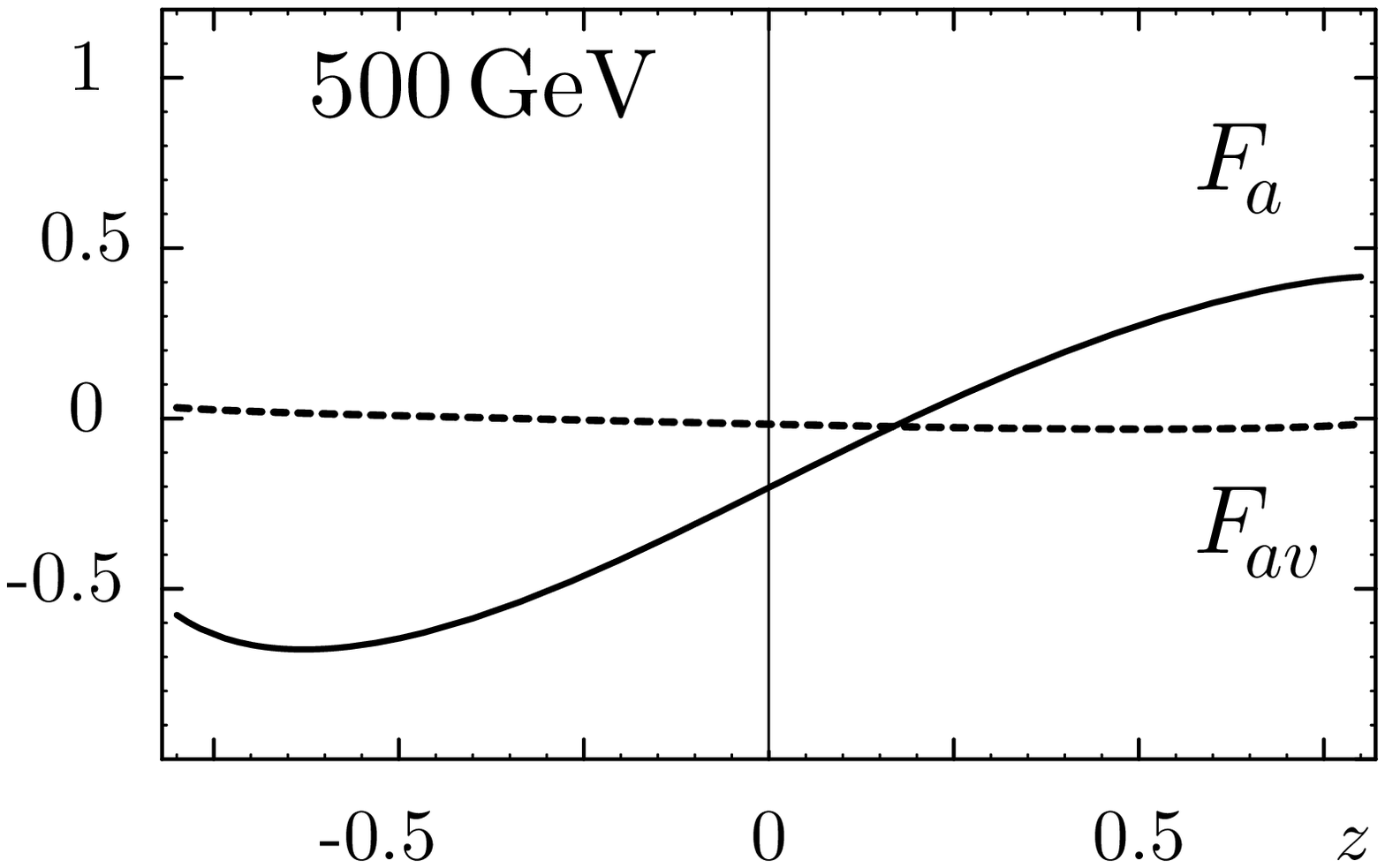,width=6cm}}
\vspace*{8pt} \caption{Factors $F_{a}(\sqrt{s},z)$ (solid) and
$F_{av}(\sqrt{s},z)$ (dashed) in the normalized deviation of the
differential cross-section $d\tilde\sigma/dz$ for $\sqrt{s}=200$
and 500 GeV.}\label{fig:5f}
\end{figure}

\subsection{One-parameter fit}

The factor $f^{ee}_2(z)$ is positive monotonic function of $z$
(see Fig. \ref{fig:0} for the center-of-mass energies
$\sqrt{s}=200$ GeV. The same behavior is observed for higher
energies). Such a property allows one to choose $f^{ee}_2(z)$ as a
normalization factor for the differential cross section. Then the
normalized deviation of the differential cross-section reads
\cite{Gulov:2004sg}
\begin{eqnarray}\label{ncs}
\frac{d\tilde\sigma}{dz}&=& \frac{m_Z^2}{{4\pi}f^{ee}_2(z)}
\Delta\,\frac{d\sigma}{dz} =
 \bar{v}^2 +
F_{a}(\sqrt{s},z) \bar{a}^2 + F_{av}(\sqrt{s},z)\bar{a}\bar{v}
+\ldots ,
\end{eqnarray}
and the normalized factors are shown in Fig \ref{fig:5f} for energies of LEP and ILC experiments. Now
these factors are finite at $z\to 1$. Each of them in a special
way influences the differential cross-section.
\begin{enumerate}
\item
The factor at $\bar{v}^2$ is just the unity. Hence, the
four-fermion contact coupling between vector currents,
$\bar{v}^2$, determines the level of the deviation from the SM
value.
\item
The factor at $\bar{a}^2$ depends on the scattering angle in a
non-trivial way. It allows  to recognize the Abelian $Z'$ boson,
if the experimental accuracy is sufficient.
\item
The factor at $\bar{a}\bar{ v}$ results in small corrections.
\end{enumerate}

Thus, effectively, the obtained normalized differential
cross-section is a two-parametric function. In the next sections
we introduce the observables to fit separately each of these
parameters.

\subsection{Observables to pick out $\bar{v}^2$}

The normalized deviation of the differential cross-section
(\ref{ncs}) is (effectively) the function of two parameters,
$\bar{a}^2$ and $\bar{v}^2$. We are going to introduce the
integrated observables which determine  separately the
four-fermion couplings $\bar{a}^2$ and $\bar{v}^2$
\cite{Gulov:2004sg}.

Let us first proceed with the observable for $\bar{v}^2$. After
normalization the factor at the vector-vector four-fermion
coupling becomes the unity. Whereas the factor at $\bar{a}^2$ is a
sign-varying function of the cosine of the scattering angle. As it
follows from Fig. \ref{fig:5f}, for the center-of-mass energy 200
GeV it is small over the backward scattering angles. So, to
measure the value of $\bar{v}^2$ the normalized deviation of the
differential cross-section has to be integrated over the backward
angles. For the center-of-mass energy 500 GeV the factor at
$\bar{a}^2$ is already a non-vanishing quantity for the backward
scattering angles. The curves corresponding to intermediate
energies are distributed in between two these curves. Since they
are sign-varying ones at each energy point some interval of $z$
can be chosen to make the integral to be zero. Thus, to measure
the $Z'$ coupling to the electron vector current $\bar{v}^2$ we
introduce the integrated cross-section (\ref{ncs})
\begin{equation}\label{vobs}
\sigma_V = \int_{z_0}^{z_0+\Delta z} (d\tilde\sigma/dz)dz,
\end{equation}
where at each energy the most effective interval $[z_0,z_0+\Delta
z]$ is determined by the following requirements:
\begin{enumerate}
\item The relative contribution of the coupling $\bar{v}^2$ is
maximal. Equivalently, the contribution of the factor at
$\bar{a}^2$ is suppressed. \item The length $\Delta z$ of the
interval is maximal. This condition ensures that the largest
number of bins is taken into consideration.
\end{enumerate}

\begin{figure}
\centerline{\psfig{file=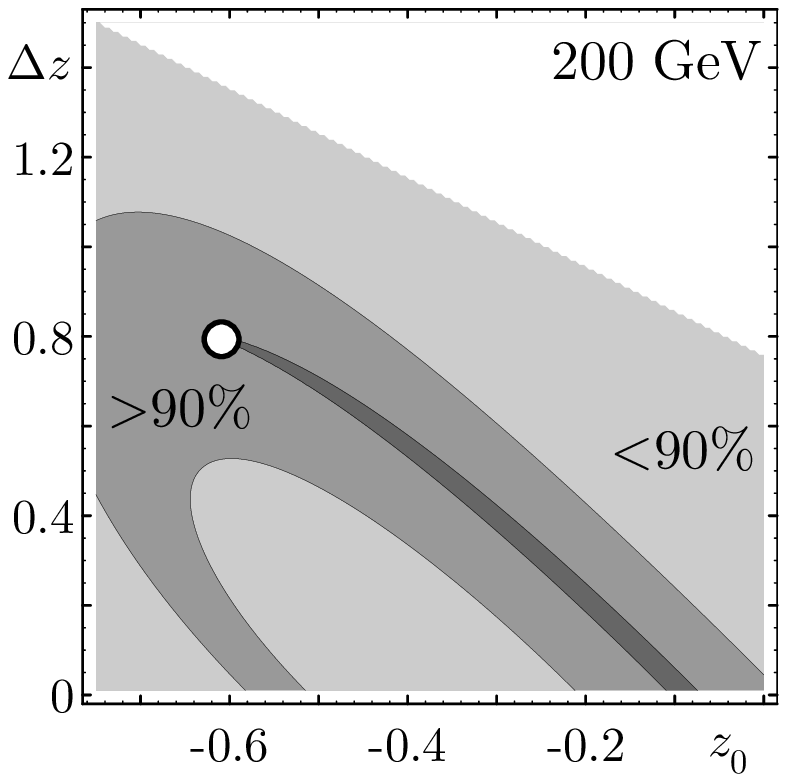,width=55mm}\rule{5mm}{0pt}\psfig{file=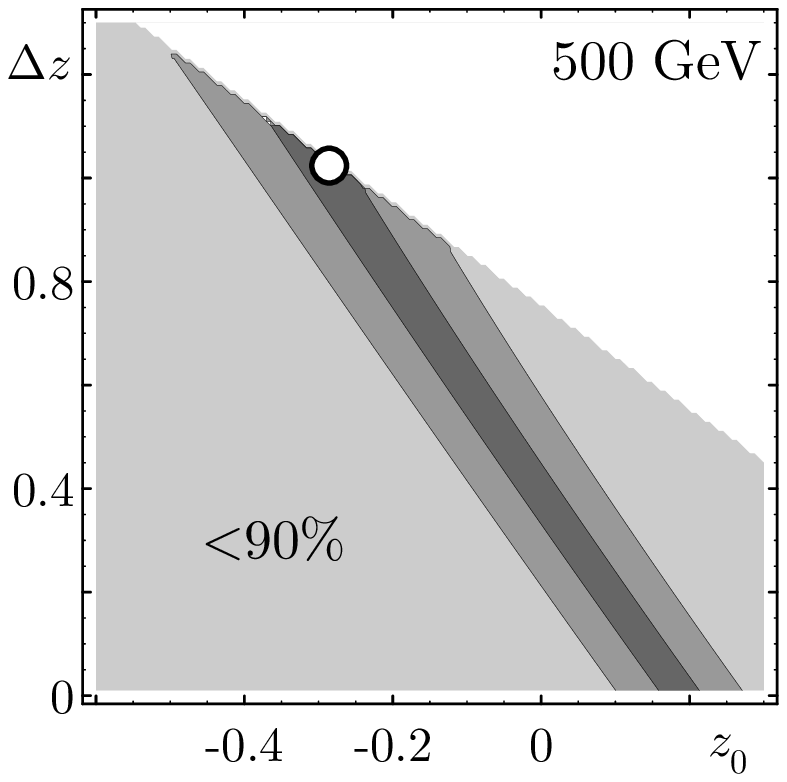,width=55mm}}
\vspace*{8pt} \caption{Relative contribution of the factor at
$\bar{v}^2$ to the observable $\sigma_V$ as the function of the
left boundary of the angle interval, $z_0$, and the interval
length, $\Delta z$, at the center-of-mass energy 200 and 500 GeV.
The shaded areas correspond to the contributions $>95\%$ (dark),
from 90\% to 95\% (midtone), and $<90\%$ (light).}\label{fig:f-6}
\end{figure}

The relative contribution of the factor at $\bar{v}^2$ is defined
as
\begin{equation}
\kappa_V= \frac{\Delta z}{\Delta z+\left|\int_{z_0}^{z_0+\Delta
z}F_a \,dz\right|+\left|\int_{z_0}^{z_0+\Delta
z}F_{av}\,dz\right|}
\end{equation}
and shown in Fig. \ref{fig:f-6} as the function of the left
boundary of the angle interval, $z_0$, and the interval length,
$\Delta z$. In each plot the dark area corresponds to the
observables which values are determined by the vector-vector
coupling $\bar{v}^2$ with the accuracy $>95\%$. The area reflects
the correlation of the width of the integration interval $\Delta
z$ with the choice of the initial $z_0$ following from the
mentioned requirements. Within this area we choose the observable
which includes the largest number of bins (largest $\Delta z$).
The corresponding values of $z_0$ and $\Delta z$ are marked by the
white dot on the plots in Fig. \ref{fig:f-6}. As the carried out
analysis showed, the point $z_0$ is shifted to the right with
increase in energy whereas $\Delta z$ remains approximately the
same.

From the plots it follows that the most efficient intervals are
\begin{eqnarray}
&& -0.6<z<0.2,\quad\sqrt{s}=200\mbox{ GeV}, \nonumber\\&&
-0.3<z<0.7,\quad\sqrt{s}=500\mbox{ GeV}.
\end{eqnarray}
Therefore the observable (\ref{vobs}) allows to measure the $Z'$
coupling to the electron vector current $\bar{v}^2$ with the
efficiency $>95\%$.

Fitting the LEP2 final data with the one-parameter observable, we
find the values of the $Z'$ coupling to the electron vector
current together with their 1$\sigma$ uncertainties:
\begin{eqnarray}
  \mathrm{ALEPH}: &\bar{v}_e^2=& -0.11\pm 6.53 \times 10^{-4}\nonumber\\
  \mathrm{DELPHI}: &\bar{v}_e^2=& 1.60\pm 1.46 \times 10^{-4} \nonumber\\
  \mathrm{L3}: &\bar{v}_e^2=& 5.42\pm 3.72 \times 10^{-4} \nonumber\\
  \mathrm{OPAL}: &\bar{v}_e^2=& 2.42\pm 1.27 \times 10^{-4} \nonumber\\
  \mathrm{Combined}: &\bar{v}_e^2=& 2.24\pm 0.92 \times 10^{-4}. \nonumber
\end{eqnarray}
As one can see, the most precise data of DELPHI and OPAL
collaborations are resulted in the Abelian $Z'$ hints at one and
two standard deviation level, correspondingly. The combined value
shows the 2$\sigma$ hint which corresponds to $0.006\le
|\bar{v}_e|\le 0.020$.

\subsection{Observables to pick out $\bar{a}^2$}

In order to pick the axial-vector coupling $\bar{a}^2$ one needs
to eliminate the dominant contribution coming from $\bar{v}^2$.
Since the factor at $\bar{v}^2$ in the $d\tilde\sigma/dz$  equals
unity, this can be done by summing up equal number of bins with
positive and negative weights. In particular, the forward-backward
normalized deviation of the differential cross-section appears to
be sensitive mainly to $\bar{a}^2$,
\begin{eqnarray}
\tilde\sigma_{\rm FB}&=&\int\nolimits_{0}^{z_{\rm
max}}dz\,\frac{d\tilde\sigma}{dz} -\int\nolimits_{-z_{\rm
max}}^{0}dz\,\frac{d\tilde\sigma}{dz}
\tilde{F}_{a,\rm FB} \bar{a}^2 + \tilde{F}_{av,\rm
FB}\bar{a}\bar{v}.
\end{eqnarray}
The value $ z_{\rm max}$ is determined by the number of bins
included and, in fact, depends on the data set considered. The
LEP2 experiment accepted $e^+e^-$ events with $|z|<0.72$. In what
follows we take the angular cut $z_{\mathrm{max}}=0.7$ for
definiteness.

The efficiency of the observable is determined as:
\begin{equation}
\kappa= \frac{|\tilde{F}_{a,\rm FB}|}{|\tilde{F}_{a,\rm
FB}|+|\tilde{F}_{av,\rm FB}|}.
\end{equation}
It can be estimated as $\kappa=0.9028$ for the center-of-mass
energy 200 GeV and $\kappa=0.9587$ for 500 GeV. Thus, the
observable
\begin{eqnarray}\label{aobs}
&& \tilde\sigma_{\rm FB}=0.224 \bar{a}^2 - 0.024
\bar{a}\bar{v},\quad\sqrt{s}=200\mbox{ GeV}, \nonumber\\&&
\tilde\sigma_{\rm FB}=0.472 \bar{a}^2 - 0.020
\bar{a}\bar{v},\quad\sqrt{s}=500\mbox{ GeV}
\end{eqnarray}
is mainly sensitive to the $Z'$ coupling to the axial-vector
current $\bar{a}^2$.

Consider a usual situation when experiment is not able to
recognize the angular dependence of the differential cross-section
deviation from its SM value with the proper accuracy because of
loss of statistics. Nevertheless, a unique signal of the Abelian
$Z'$ boson can be determined. For this purpose the observables
$\int_{z_0}^{z_0+\Delta z}(d\tilde\sigma/dz)dz$ and
$\tilde\sigma_{\rm FB}$ must be measured. Actually, they are
derived from the normalized deviation of the differential
cross-section. If the deviation is inspired by the Abelian $Z'$
boson both the observables are to be positive quantities
simultaneously. This feature serves as the distinguishable signal
of the Abelian $Z'$ virtual state in the Bhabha process for the
LEP2 energies as well as for the energies of future
electron-positron collider ILC ($\geq 500$ GeV). The observables
fix the unknown low energy vector and axial-vector $Z'$ couplings
to the electron current. Their values have to be correlated with
the bounds on $\bar{a}^2$ and $\bar{v}^2$ derived by means of
independent fits for other scattering processes.

We estimated the observable (\ref{aobs}) related to the value of
$\bar{a}^2$. Since in the Bhabha process the effects of the
axial-vector coupling are suppressed with respect to those of the
vector coupling, we expect much larger experimental  uncertainties
for $\bar{a}^2$. Indeed, the LEP2 data lead to the huge errors for
$\bar{a}^2$ of order $10^{-3}-10^{-4}$. The mean values are
negative numbers which are too large to be interpreted as a
manifestation of some heavy virtual state beyond the energy scale
of the SM.

Thus, the LEP2 data constrain the value of $\bar{v}^2$ at the
$2\sigma$ CL which could correspond to the Abelian $Z'$ boson with
the mass of the order 1~TeV. In contrast, the value of $\bar{a}^2$
is a large negative number with a significant experimental
uncertainty. This can not be interpreted as a manifestation of
some heavy virtual state beyond the energy scale of the SM.

\subsection{Many-parameter fits}
To account for all the accessible data of LEP experiments, we
address to many parameter fits \cite{Gulov:2007zz}. As the basic
observable to fit the LEP2 experiment data on the Bhabha process
we propose the differential cross-section
\begin{equation}\label{7}
\left.\frac{d\sigma^\mathrm{Bhabha}}{dz}-\frac{d\sigma^{\mathrm{Bhabha},SM}}{dz}\right|_{z=z_i,\sqrt{s}=\sqrt{s_i}},
\end{equation}
where $i$ runs over the bins at various center-of-mass energies
$\sqrt{s}$. The final differential cross-sections measured by the
ALEPH \cite{Barate:1999qx} (130-183 GeV), DELPHI
\cite{Abdallah:2005ph} (189-207 GeV), L3 \cite{Acciarri:1999rw}
(183-189 GeV), and OPAL
\cite{Abbiendi:2003dh,Abbiendi:1998ea,Ackerstaff:1997nf} (130-207
GeV) collaborations are taken into consideration (299 bins).

As the observables for $e^+e^-\to\mu^+\mu^-,\tau^+\tau^-$
processes, we consider the total cross-section and the
forward-backward asymmetry
\begin{equation}\label{8}
\sigma^{\ell^+\ell^-}_T-\sigma_T^{\ell^+\ell^-,\mathrm{SM}},
\quad
\left.A^{\ell^+\ell^-}_{FB}-A_{FB}^{\ell^+\ell^-,\mathrm{SM}}\right|_{\sqrt{s}=\sqrt{s_i}},
\end{equation}
where $i$ runs over 12 center-of-mass energies $\sqrt{s}$ from 130
to 207 GeV. We consider the combined LEP2 data
\cite{Alcaraz:2006mx} for these observables (24 data entries for
each process). These data are more precise as the corresponding
differential cross-sections. Our analysis is based on the fact
that the kinematics of $s$-channel processes is rather simple and
the differential cross-section is effectively a two-parametric
function of the scattering angle. The total cross-section and the
forward-backward asymmetry incorporate complete information about
the kinematics of the process and therefore are an adequate
alternative for the differential cross-sections.

The data are analysed by means of the $\chi^2$ fit
\cite{Gulov:2007zz}. Denoting the observables (\ref{7})--(\ref{8})
by $\sigma_i$, one can construct the $\chi^2$-function,
\begin{equation}\label{9}
\chi^2(\bar{a}, \bar{v}_e,\bar{v}_\mu,\bar{v}_\tau) =
\sum\limits_i
\left[\frac{\sigma^\mathrm{ex}_i-\sigma^\mathrm{th}_i(\bar{a},
\bar{v}_e,\bar{v}_\mu,\bar{v}_\tau)}{\delta\sigma_i}\right]^{2},
\end{equation}
where $\sigma^\mathrm{ex}$ and $\delta\sigma$ are the experimental
values and the uncertainties of the observables, and
$\sigma^\mathrm{th}$ are their theoretical expressions presented
in Eqs. (\ref{4})--(\ref{5}). The sum in Eq. (\ref{9}) refers to
either the data for one specific process or the combined data for
several processes. By minimizing the $\chi^2$-function, the
maximal-likelihood estimate for the $Z'$ couplings can be derived.
The $\chi^2$-function is also used to plot the confidence area in
the space of parameters $\bar{a}$, $\bar{v}_e$, $\bar{v}_\mu$, and
$\bar{v}_\tau$. Note that in this way of experimental data
treating all the possible correlations are neglected. We believe
that at the present stage of investigation this is reasonable,
because the Collaborations have never reported on this
possibility.

For all the considered processes, the theoretic predictions
$\sigma^\mathrm{th}_i$ are linear combinations of products of two
$Z'$ couplings
\begin{equation}\label{10}
\sigma^\mathrm{th}_i=\sum_{j=1}^{7} C_{ij}A_j,\qquad
 A_j=\{\bar{a}^2,\bar{v}_e^2,\bar{a}\bar{v}_e,\bar{v}_e\bar{v}_\mu,\bar{v}_e\bar{v}_\tau,
\bar{a}\bar{v}_\mu,\bar{a}\bar{v}_\tau\}, \nonumber
\end{equation}
where $C_{ij}$ are known numbers.

In the Bhabha process, the $Z'$ effects are determined by three
linear-independent contributions coming from $\bar{a}^2$,
$\bar{v}_e^2$, and $\bar{a}\bar{v}_e$ and the number degrees of
freedom (d.o.f) $M=3$. As for the
$e^+e^-\to\mu^+\mu^-,\tau^+\tau^-$ processes, the observables
depend on four linear-independent terms for each process:
$\bar{a}^2$, $\bar{v}_e\bar{v}_\mu$, $\bar{v}_e\bar{a}$,
$\bar{a}\bar{v}_\mu$ for $e^+e^-\to\mu^+\mu^-$; and $\bar{a}^2$,
$\bar{v}_e\bar{v}_\tau$, $\bar{v}_e\bar{a}$, $\bar{a}\bar{v}_\tau$
for $e^+e^-\to\tau^+\tau^-$ ($M=4$). Note that some terms in the
observables for different processes are the same. Therefore, the
number of $\chi^2$ d.o.f. in the combined fits is less than the
sum of d.o.f. for separate processes. Hence, the predictive power
of the larger set of data is not drastically spoiled by the
increased number of d.o.f. In fact, combining the data of the
Bhabha and $e^+e^-\to\mu^+\mu^-$ ($\tau^+\tau^-$) processes
together we have to treat five linear-independent terms. The
complete data set for all the lepton processes is ruled by seven
d.o.f. As a consequence, the combination of the data for all the
lepton processes is possible.

The parametric space of couplings ($\bar{a}$, $\bar{v}_e$,
$\bar{v}_\mu$, $\bar{v}_\tau$) is four-dimensional. However, for
the Bhabha process it is reduced to the plane ($\bar{a}$,
$\bar{v}_e$), and to the three-dimensional volumes ($\bar{a}$,
$\bar{v}_e$, $\bar{v}_\mu$), ($\bar{a}$, $\bar{v}_e$,
$\bar{v}_\tau$) for the $e^+e^-\to\mu^+\mu^-$ and
$e^+e^-\to\tau^+\tau^-$ processes, correspondingly. The predictive
power of data is distributed not uniformly over the parameters.
The parameters $\bar{a}$ and $\bar{v}_e$ are present in all the
considered processes and appear to be significantly constrained.
The couplings $\bar{v}_\mu$ or $\bar{v}_\tau$ enter when the
processes $e^+e^-\to\mu^+\mu^-$ or $e^+e^-\to\tau^+\tau^-$ are
accounted for. So, in these processes, we also study the
projection of the confidence area onto the plane
($\bar{a},\bar{v}_e$).

The origin of the parametric space, $\bar{a}=\bar{v}_e=0$,
corresponds to the absence of the $Z'$ signal. This is the SM
value of the observables. This point could occur inside or outside
of the confidence area at a fixed CL. When it lays out of the
confidence area, this means the distinct signal of the Abelian
$Z'$. Then the signal probability can be defined as the
probability that the data agree with the Abelian $Z'$ boson
existence and exclude the SM value. This probability corresponds
to the most stringent CL (the largest $\chi^2_\mathrm{CL}$) at
which the point $\bar{a}=\bar{v}_e=0$ is excluded. If the SM value
is inside the confidence area, the $Z'$ boson is indistinguishable
from the SM. In this case, upper bounds on the $Z'$ couplings can
be determined.

\begin{figure}
\begin{minipage}{0.45\linewidth}
\centerline{\psfig{file=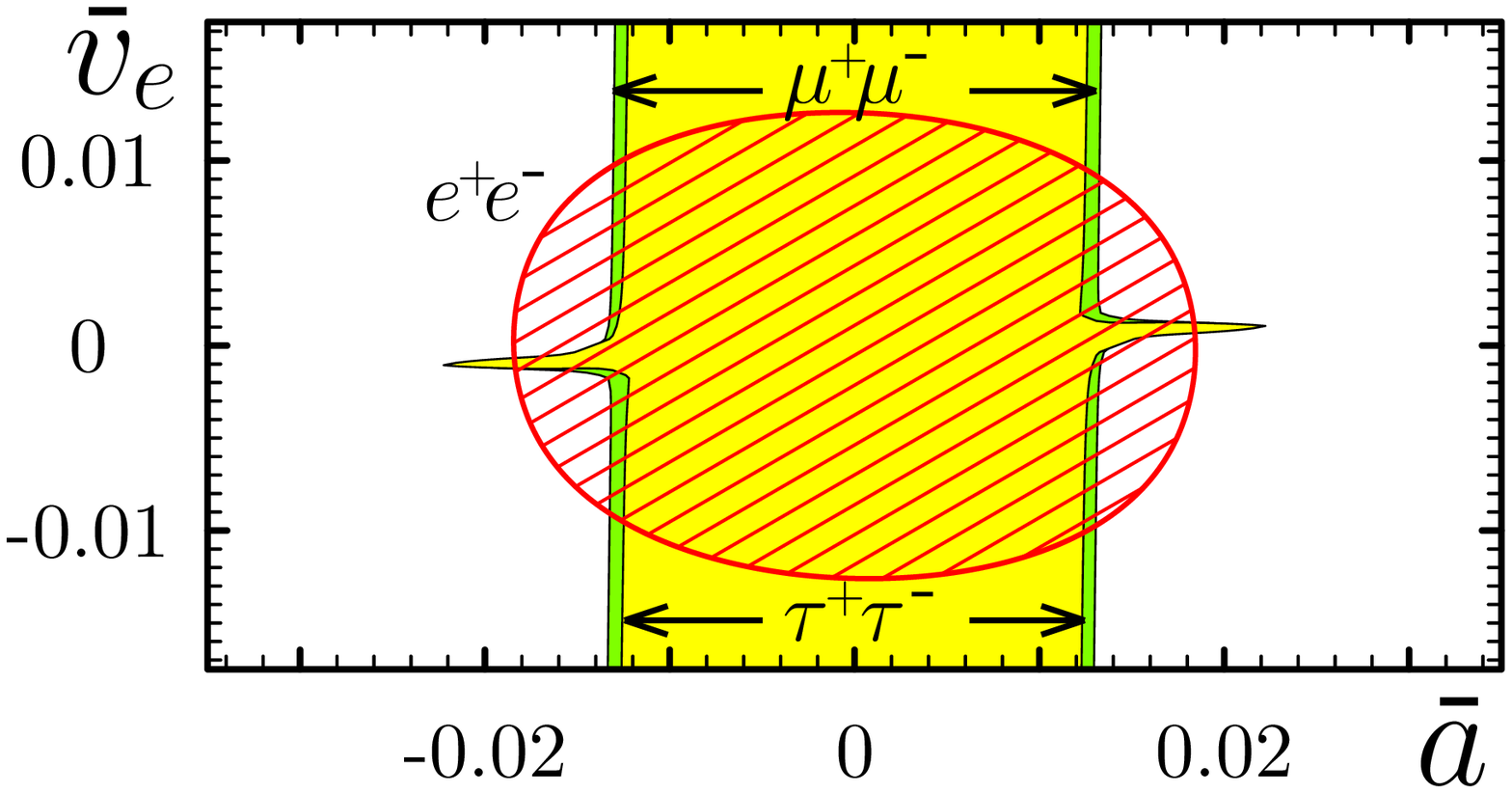,width=60mm}}
  \caption{The 95\% CL areas in the ($\bar{a},\bar{v}_e$) plane for the Bhabha,
  $e^+e^-\to\mu^+\mu^-$, and $e^+e^-\to\tau^+\tau^-$ processes.}\label{fig:1}
\end{minipage}
\hfil
\begin{minipage}{0.45\linewidth}
\centerline{\psfig{file=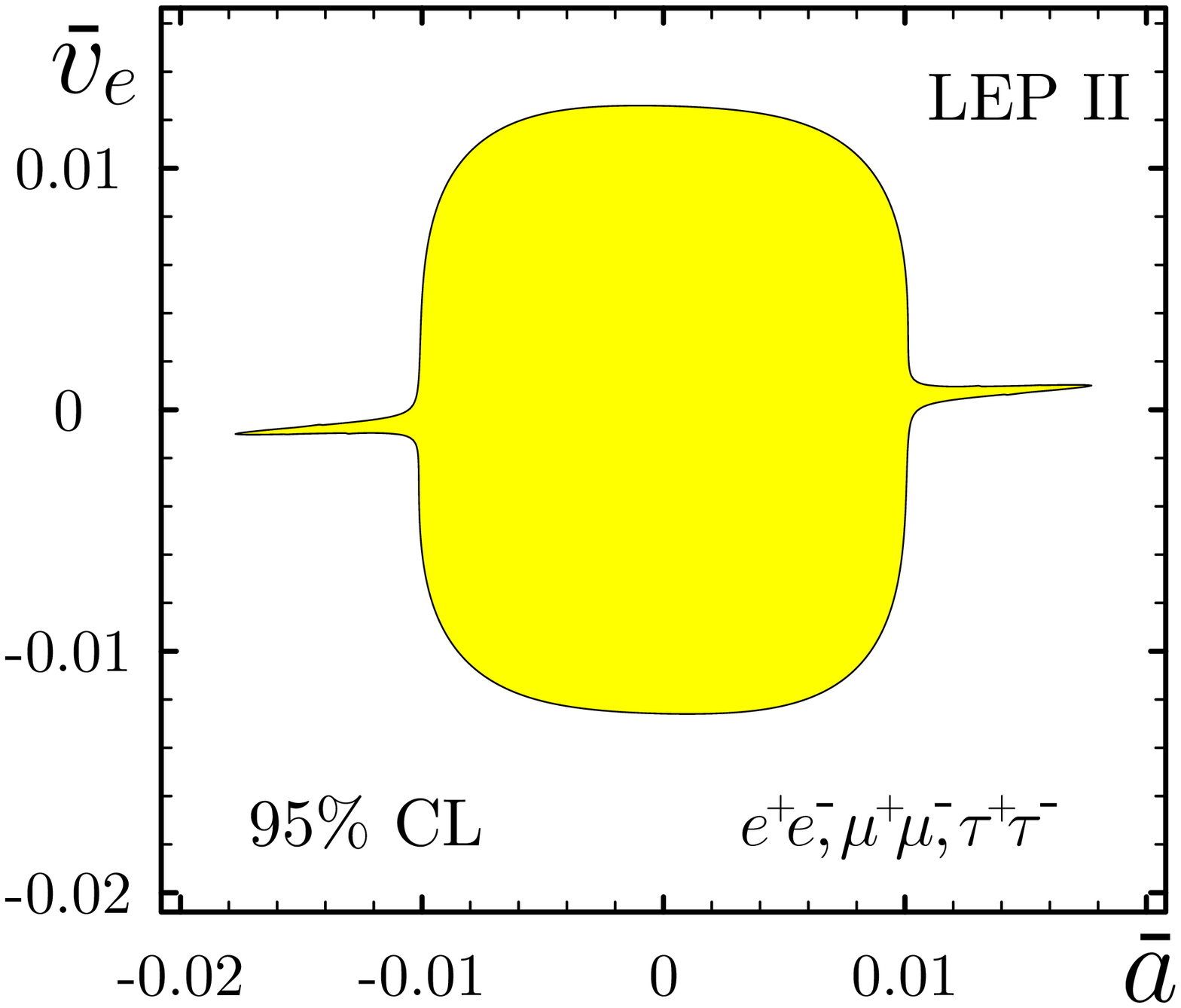,width=60mm}}
  \caption{The projection of the 95\% CL area onto the ($\bar{a},\bar{v}_e$) plane
  for the combination of the Bhabha, $e^+e^-\to\mu^+\mu^-$, and $e^+e^-\to\tau^+\tau^-$
  processes.}\label{fig:2}
\end{minipage}
\end{figure}

The 95\% CL areas in the ($\bar{a},\bar{v}_e$) plane for the
separate processes are plotted in Fig. \ref{fig:1}. As it is seen,
the Bhabha process constrains both the axial-vector and vector
couplings. As for the $e^+e^-\to\mu^+\mu^-$ and
$e^+e^-\to\tau^+\tau^-$ processes, the axial-vector coupling is
significantly constrained, only. The confidence areas include the
SM point at the meaningful CLs, so the experiment could not pick
out clearly the Abelian $Z'$ signal from the SM. An important
conclusion from these plots is that the experiment significantly
constrains only the couplings entering sign-definite terms in the
cross-sections.

The combination of all the lepton processes is presented in Fig.
\ref{fig:2}. There is no visible signal beyond the SM. The
couplings to the vector and axial-vector electron currents are
constrained by the many-parameter fit as $|\bar{v}_e|<0.013$,
$|\bar{a}|<0.019$ at the 95\% CL. If the charge corresponding to
the $Z'$ interactions is assumed to be of order of the
electromagnetic one, then the $Z'$ mass should be greater than
0.67 TeV. For the charge of order of the SM $SU(2)_L$ coupling
constant $m_{Z'}\ge 1.4$ TeV. One can see that the constraint is
not too severe to exclude the $Z'$ searches at the LHC.

\begin{figure}
\centerline{\psfig{file=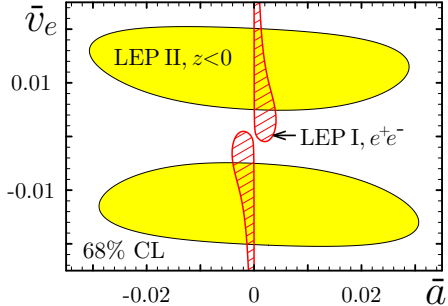,width=60mm}} \caption{The 68\% CL
area in the ($\bar{a},\bar{v}_e$) plane from the backward bins of
the Bhabha process in the LEP2 experiments (the shaded area). The
hatched area is the 68\% CL area from the LEP 1 data on the Bhabha
process.}\label{fig:3}
\end{figure}

Let us compare the obtained results with the one-parameter fits.
As one can see, the most precise data of DELPHI and OPAL
collaborations are resulted in the Abelian $Z'$ hints at one and
two standard deviation level, correspondingly. The combined value
shows the 2$\sigma$ hint, which corresponds to $0.006\le
|\bar{v}_e|\le 0.020$. On the other hand, our many-parameter fit
constrains the $Z'$ coupling to the electron vector current as
$|\bar{v}_e|\le 0.013$ with no evident signal. Why does the
one-parameter fit of the Bhabha process show the 2$\sigma$ CL hint
whereas there is no signal in the two-parameter one? Our
one-parameter observable accounts mainly for the backward bins.
This is in accordance with the kinematic features of the process:
the backward bins depend mainly on the vector coupling
$\bar{v}^2_e$, whereas the contributions of other couplings are
kinematically suppressed (see Fig. \ref{fig:0}). Therefore, the
difference of the results can be inspired by the data sets used.
To clarify this point, we perform the many-parameter fit with the
113 backward bins ($z\le 0$), only. The $\chi^2$ minimum,
$\chi^2_\mathrm{min}=93.0$, is found in the non-zero point
$|\bar{a}|=0.0005$, $\bar{v}_e= 0.015$. This value of the $Z'$
coupling $\bar{v}_e$ is in an excellent agreement with the mean
value obtained in the one-parameter fit. The 68\% confidence area
in the ($\bar{a},\bar{v}_e$) plane is plotted in Fig. \ref{fig:3}.
There is a visible hint of the Abelian $Z'$ boson. The zero point
$\bar{a}=\bar{v}_e=0$ (the absence of the $Z'$ boson) corresponds
to $\chi^2=97.7$. It is covered by the confidence area with
$1.3\sigma$ CL. Thus, the backward bins show the $1.3\sigma$ hint
of the Abelian $Z'$ boson in the many-parameter fit. So, the
many-parameter fit is less precise than the analysis of the
one-parameter observables.

\begin{table}
\caption{The summary of the fits of the LEP data for the
dimensionless contact couplings (\ref{6}).}
{\begin{tabular}{@{}ccc@{}}
  \hline
  Data & $\bar{v}^2_e$, $\times 10^{-4}$ & $\bar{a}^2$, $\times 10^{-5}$ \\
  \hline
  \multicolumn{3}{c}{LEP1} \\
  \hline
  $e^-e^+$, 68\% CL & - & $1.25\pm1.25$ 
  \\
  \hline
  \multicolumn{3}{c}{LEP2, one-parameter fits} \\
  \hline
  $e^-e^+$, 68\% CL & $2.24\pm 0.92$ 
  & - \\
  $\mu\mu$, 68\% CL & - & $3.66^{+4.89}_{-4.86}$ 
  \\
  $\mu\mu$,$\tau\tau$, 68\% CL & - & $1.33^{+3.89}_{-3.87}$\\
  \hline
  \multicolumn{3}{c}{LEP2, many-parameter fits} \\
  \hline
  $e^-e^+,\mu\mu,\tau\tau$, 95\% CL & $\le 1.69$ & $\le 36.1$\\
  $e^-e^+$ backward, 68\% CL & $2.25^{+1.79}_{-2.07}$ & $\le 94.9$\\
  \hline
\end{tabular}\label{tab:sum:1}}
\end{table}

At LEP1 experiments \cite{LEP1:2005ema} the $Z$-boson couplings to
the vector and axial-vector lepton currents ($g_V$, $g_A$) were
precisely measured. The Bhabha process shows the 1$\sigma$
deviation from the SM values for Higgs boson masses $m_H\ge 114$
GeV (see Fig. 7.3 of Ref. \cite{LEP1:2005ema}). This deviation
could be considered as the effect of the $Z$--$Z'$ mixing. It is
interesting to estimate the bounds on the $Z'$ couplings following
from these experiments.

Due to relations (\ref{grgav}), the $Z$--$Z'$ mixing angle is
completely determined by the axial-vector coupling $\bar{a}$. So,
the deviations of $g_V$, $g_A$ from their SM values are governed
by the couplings $\bar{a}$ and $\bar{v}_e$,
\begin{equation}\label{lep1}
g_V-g_V^{\mathrm{SM}}=-49.06 \bar{a}\bar{v}_e,\quad
g_A-g_A^{\mathrm{SM}} = 49.06 \bar{a}^2.
\end{equation}
Let us assume that the total deviation of theory from experiments
follows due to the $Z$--$Z'$ mixing. This gives an upper bound on
the $Z'$ couplings. In this way one can estimate whether  the $Z'$
boson is excluded by the experiments or not.

The 1$\sigma$ CL area for the Bhabha process from Ref.
\cite{LEP1:2005ema} is converted into the ($\bar{a},\bar{v}_e$)
plane in Fig. \ref{fig:3}. The SM values of the couplings
correspond to the top quark mass $m_t=178$ GeV and the Higgs
scalar mass $m_H=114$ GeV. As it is seen, the LEP1 data on the
Bhabha process is compatible with the Abelian $Z'$ existence at
the $1\sigma$ CL. The axial-vector coupling is constrained as
$|\bar{a}|\le 0.005$. This bound corresponds to $\bar{a}^2\le
2.5\times 10^{-5}$, which agrees with the one-parameter fits of
the LEP2 data for $e^+e^-\to\mu^+\mu^-,\tau^+\tau^-$ processes
($\bar{a}^2= 1.3\pm 3.89\times 10^{-5}$ at 68\% CL). On the other
hand, the vector coupling constant $\bar{v}_e$ is practically
unconstrained by the LEP1 experiments.

For the convenience, in Table \ref{tab:sum:1} we collect the
summary of the fits of the LEP data in terms of dimensionless
contact couplings (\ref{6}). From the analysis carried out we come
to conclusion that, in principle, the LEP experiments were able to
detect the $Z'$-boson signals if the statistics had been
sufficient.

\section{Model independent results and search for $Z'$ at the LHC}

In this section we  discuss  all the assumptions giving a
possibility to pick out the $Z'$ signal and determine its
characteristics in a model independent way. We also note the role
of the present results for the  LHC and future ILC experiments.

As it was already noted, in searching for this particle at the LEP
and Tevatron a model dependent analysis was applied. As the main
motivation for this approach it was  the different number of
chiral fermions involved in different models (see, for example,
Ref. \cite{Langacker:2008yv}). In this way the low bounds on
$m_{Z'}$ have been estimated and the smallness of the $Z$--$Z'$
mixing was also observed.

On the contrary, in our approach the relations (\ref{rgr2})
between the parameters of the effective low energy Lagrangians
have been accounted for that gave a possibility to determine not
only the bounds but also the mass and other parameters of the
$Z'$.

To be precise, let us note all the assumptions used in our
investigations. We analyzed the four-fermion scattering amplitudes
of order $\sim m_{Z'}^{-2}$ generated by the $Z'$ virtual states.
The vertices linear in $Z'$ were included into the effective
low-energy Lagrangian. We also impose a number of natural
conditions. The interactions of a renormalizable type are dominant
at low energies $\sim m_W$. The non-renormalizable interactions
generated at high energies due to radiation corrections are
suppressed by the inverse heavy mass $\sim 1/m_{Z'}$ and
neglected. We also assumed that the $SU(2)_L \times U(1)_Y$ gauge
group of the SM is a subgroup of the GUT group. As a consequence,
all the structure constants connecting two SM gauge bosons with
$Z'$ have to be zero. Hence, the interactions of gauge fields of
the types $Z' W^+ W^-, Z' Z Z$, and other are absent at a tree
level. Our effective Lagrangian is also consistent with the
absence  of the tree-level flavor-changing neutral currents
(FCNCs) in the fermion sector. The renormalizable interactions of
fermions and scalars are described by the Yukawa Lagrangian that
accounts for different possibilities of the Yukawa sector without
the tree-level FCNCs. These assumptions are quite general and
satisfied in a wide class of $E_6$ inspired models.

Within these constraints for the low energy effective Lagrangian
the relations (\ref{rgr2}),(\ref{grgav}) have been derived.
Correspondingly, the model independent estimates of the mass
$m_{Z'}$ and other parameters are regulated by  the noted
requirements.  Therefore, the extended underlying model has also
to accept  them.

In this regard, let us discuss the role of the obtained estimates
for the LHC. As it is well known (see, for example,
\cite{Langacker:2008yv,Rizzo:2006nw}), there are many tools at the
LHC for $Z'$ identification. But many of them are only applicable
if this particle is relatively light. Our results are in favor to
this case.

Next important point is the determination of $Z'$ couplings to the
various SM fermions. As we have shown, the axial-vector couplings
of the $Z'$ to the SM fermions are universal and proportional to
its coupling to the Higgs field. Hence we have obtained an
estimate of the $a = a_f$ couplings for both leptons and quarks.
This is an essential input because experimental analysis  for the
LHC have mainly concentrated on being able to distinguish models
and not on actual couplings.\footnote{Discussion of the
determination of couplings can be found in Ref.
\cite{Petriello:2008zr}.} The vector coupling $v_e$ was also
estimated that, in particular, may help to distinguish the decay
of the $Z'$ resonance state to $ e^+ e^-$ pairs. Since the
couplings $a_e$ and $v_e$ were estimated there is a possibility to
distinguish this process from the decay of the $K K $ system. In
the literature on searching for the $Z'$ it is also mentioned
\cite{Rizzo:2006nw,Dittmar:2003ir} that the determination of the
$Z'$ couplings to fermions could be fulfilled channel by channel,
$a_e, v_e, v_{e,b}, a_{e,b}$, \ldots. In that considerations the
relations between the parameters  have not been taken into
account. But this is very essential for  treating of experimental
data and introducing  the relevant observables to measure. Our
consideration could be useful in this problem.

Other  parameter is the $Z$--$Z'$ mixing which is responsible for
the different decay processes and the effective interaction
vertices generated at the LHC
\cite{Langacker:2008yv,Rizzo:2006nw}. It is also determined by the
axial-vector coupling (see Eq. (\ref{grgav})) and estimated in a
model independent way. Remind that in our analysis  the mixing was
systematically  accounted for. Its value is of the same order of
magnitude as the parameters that were fitted in experiments. It
worth to note also that the existence of other heavy particles
with  masses $m_X \ge m_{Z'}$ does not influence the RG relations.

An important role of the model independent results for searching
for  $Z'$ at the Tevatron, LHC and ILC consists, in particular, in a
possibility to determine the particle as a virtual state due to a
large amount of relevant events. We mentioned already that, in
principle, LEP2 experiments were able to determine it if the
statistics was sufficiently large. Experiments at the ILC will
increase in many times  the data set of interest. In fact, the
observables, introduced in sects. 6 and 7 for picking out uniquely
$a_f^2$ and $v_e^2$ couplings in the leptonic scattering process,
are also effective at energies $\sqrt{s}\ge 500$ GeV and could be
applied in future experiments at ILC.

Other model independent methods of searching for the $Z'$ as a
resonance state are proposed in the literature (see Refs.
\cite{Dittmar:2003ir,Coriano:2008wf,Petriello:2008zr}). We do not
discuss them here because they take into consideration no
relations between the parameters. As we mentioned already, the
main goal of the present paper is to adduce model independent
information about the $Z'$ followed from experiments at low
energies. Different aspects of $Z'$ physics at the LHC are out of
the scope of it.

\section{Discussion}

In this section  we collect in a convenient form all the results
obtained  and make a comparison with other investigations on
searching  for $Z'$ at low energies. In fact, this is a large area
to discuss. References to numerous results obtained in either
model dependent or model independent approaches can be found in
Refs. \cite{Langacker:2008yv,Rizzo:2006nw}. Further subdivision
can be done into the considerations  accounting for any type
correlations between the parameters of the low energy effective
interactions and that of assuming complete independence of them.
Because of a large amount of fitting parameters the latter are
less predictable.

Now, for a convenience of readers we present the results of fits
of the $Z'$ parameters  in terms of the popular notations
\cite{Leike:1998wr,Langacker:2008yv}. The Lagrangian reads
\begin{eqnarray}\label{avstandard}
{\cal L}_{Z\bar{f}f}&=&\frac{1}{2} Z_\mu\bar{f}\gamma^\mu\left[
(v^{\mathrm{SM}}_f+ \Delta^V_f) - \gamma^5
(a^{\mathrm{SM}}_f+\Delta^A_f) \right]f, \nonumber\\
{\cal L}_{Z'\bar{f}f}&=&\frac{1}{2} Z'_\mu\bar{f}\gamma^\mu\left[
(v'_f-\gamma^5 a'_f)\right]f,
\end{eqnarray}
with the SM values of the $Z$ couplings
\[
v^{\mathrm{SM}}_f = \frac{e\left(T_{3f}-2Q_f
\sin^2\theta_W\right)}{\sin\theta_W\cos\theta_W},\qquad
a^{\mathrm{SM}}_f = \frac{e\,T_{3f}}{\sin\theta_W\cos\theta_W},
\]
where $e$ is the positron charge, $Q_f$ is the fermion charge in
the units of $e$, $T_{3f}=1/2$ for the neutrinos and $u$-type
quarks, and $T_{3f}=-1/2$ for the charged leptons and $d$-type
quarks.

\begin{table}
\caption{The summary of the fits of the LEP data for the maximum
likelihood values of the $Z'$ couplings (\ref{avstandard}) to the
SM fermions and of the $Z$--$Z'$ mixing angle $\theta_0$.
$M=\frac{m_{Z'}}{1\, \mathrm{TeV}}$ denotes the unknown value of
the $Z'$ mass in TeV units.}{ \label{fitMLV}
\begin{tabular}{@{}lcccc@{}}
  \hline
  Data & $|\theta_0|$, $\times 10^{-3}$ & $|v'_e|$, $\times 10^{-1}$ & $|a'_f|$, $\times 10^{-1}$ & $\Delta^A_e$, $\times 10^{-3}$ \\
  \hline
  \multicolumn{5}{c}{LEP1} \\
  \hline
  $e^-e^+$                  & ${3.17}{M}^{-1}$ & -     & $1.38M$ & 0.437 \\
  \hline
  \multicolumn{5}{c}{LEP2, one-parameter fits} \\
  \hline
  $e^-e^+$                  & -              & $5.83M$ & -        & - \\
  $\mu^-\mu^+$              & $5.42{M}^{-1}$ & -       & $2.36M$ & 1.278 \\
  $\mu^-\mu^+,\tau^-\tau^+$ & $3.27{M}^{-1}$ & -       & $1.42M$ & 0.464 \\
  \hline
  \multicolumn{5}{c}{LEP2, many-parameter fits} \\
  \hline
  $e^-e^+$, $z<0$           & -              & $5.84M$ & -        & - \\
  \hline
\end{tabular}}
\end{table}
\begin{table}
\caption{The summary of the fits of the LEP data for the
confidence intervals for the $Z'$ couplings (\ref{avstandard}) to
the SM fermions and for the $Z$--$Z'$ mixing angle $\theta_0$.
$M=\frac{m_{Z'}}{1\, \mathrm{TeV}}$ denotes the unknown value of
the $Z'$ mass in TeV units.}{ \label{fitCLI}
\begin{tabular}{@{}lccccc@{}}
  \hline
  Data & CL & $|\theta_0|$, $\times 10^{-3}$ & $|v'_e|$, $\times 10^{-1}$ & $|a'_f|$, $\times 10^{-1}$ & $\Delta^A_e$, $\times 10^{-3}$ \\
  \hline  \multicolumn{6}{c}{LEP1} \\
  \hline  $e^-e^+$ & 68\% & $(0;4.48){M}^{-1}$ & - & $(0;1.95)M$ & $(0;0.873)$ \\
  \hline  \multicolumn{6}{c}{LEP2, one-parameter fits} \\
  \hline  $e^-e^+$ & 95\% & - & $(2.46;7.87)M$ & - & - \\
  $\mu^-\mu^+$ & 95\% & $(0;10.39){M}^{-1}$ & - & $(0;4.52)M$ & $(0;4.694)$ \\
  $\mu^-\mu^+$,
  $\tau^-\tau^+$ & 95\% & $(0;8.64){M}^{-1}$ & - & $(0;3.75)M$ & $(0;3.244)$ \\
  \hline  \multicolumn{6}{c}{LEP2, many-parameter fits} \\
  \hline  $e^-e^+$,
  $\mu^-\mu^+$, $\tau^-\tau^+$ & 95\% & $(0;17.03){M}^{-1}$ & $(0;5.06)M$ & $(0;7.40)M$ & $(0;12.607)$ \\
  $e^-e^+$,
  $z<0$ & 68\% & $(0;27.61){M}^{-1}$ & $(1.68; 7.83)M$ & $(0;12.00)M$ & $(0;33.1288)$ \\
  \hline
\end{tabular}}
\end{table}
The results of fits of the $Z'$ couplings to the SM leptons
obtained from the analysis of LEP experiments are adduced in the
Tables \ref{fitMLV}-\ref{fitCLI}. Remind that due to the
universality of the axial-vector coupling $a_f$ the same estimates
also hold for quarks. First of all, one parameter fits of LEP
experiments as well as the many-parameter fit for the $e^+e^-$
backward bins show the hints of the $Z'$ boson at the 1-2$\sigma$
CL. Due to this fact, the fits allow to determine the maximum
likelihood values of $Z'$ parameters. In spite of uncertainties,
these values can be used as a guiding line for the estimation of
possible $Z'$ effects in the Tevatron and LHC experiments. The
maximum likelihood values are given in Table \ref{fitMLV}. As it
is seen, different fits obtained for different processes lead to
the comparable values of the $Z'$ parameters.

\begin{figure}
\centerline{\psfig{file=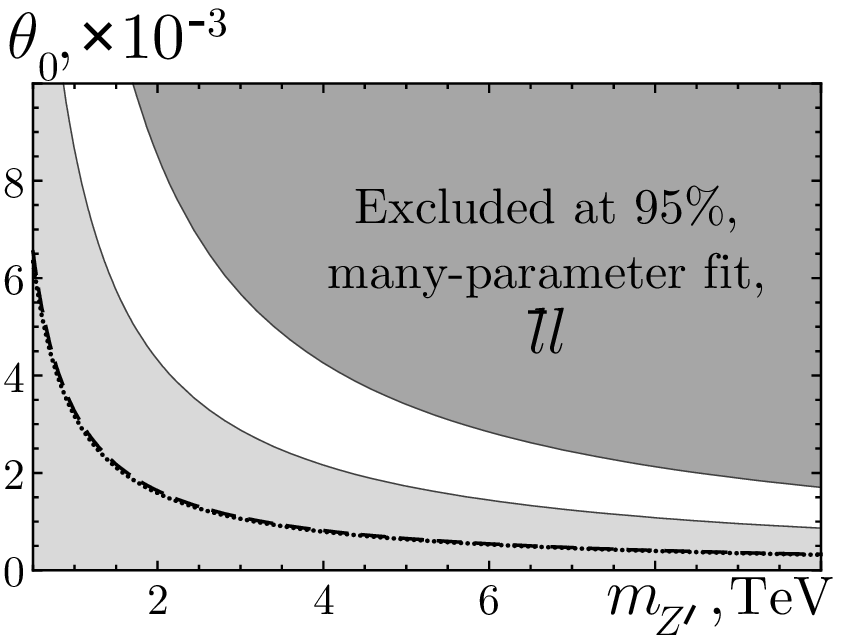,width=55mm}\hfil\psfig{file=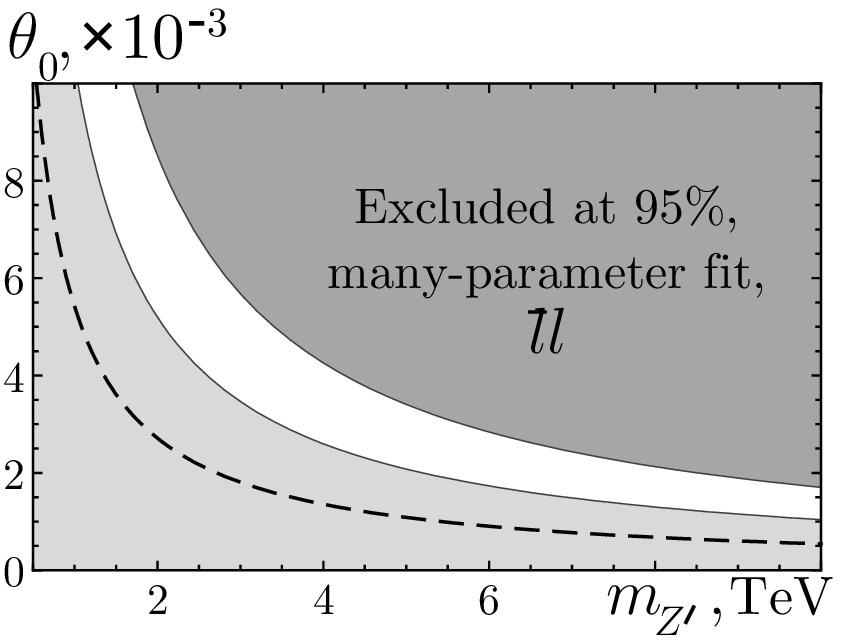,width=55mm}}

\centerline{\psfig{file=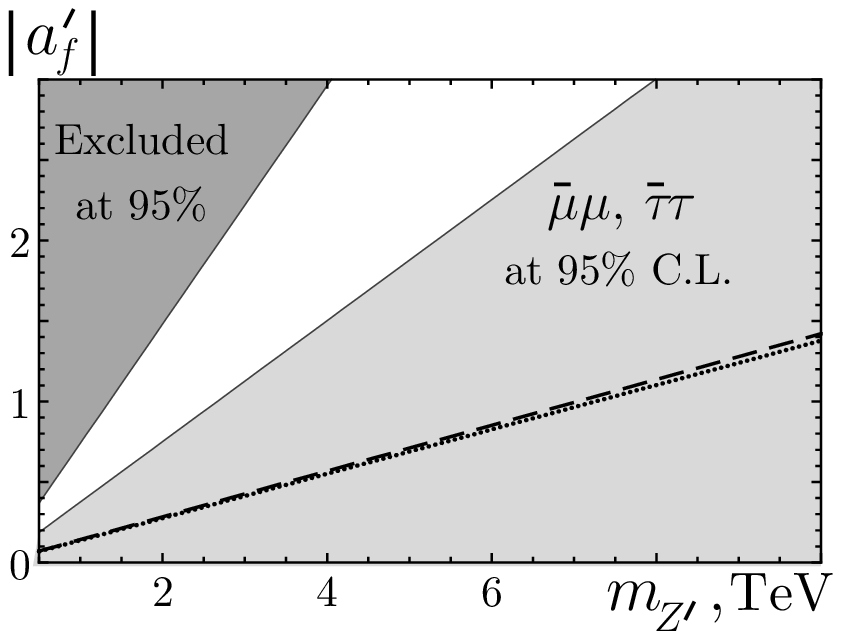,width=55mm}\hfil\psfig{file=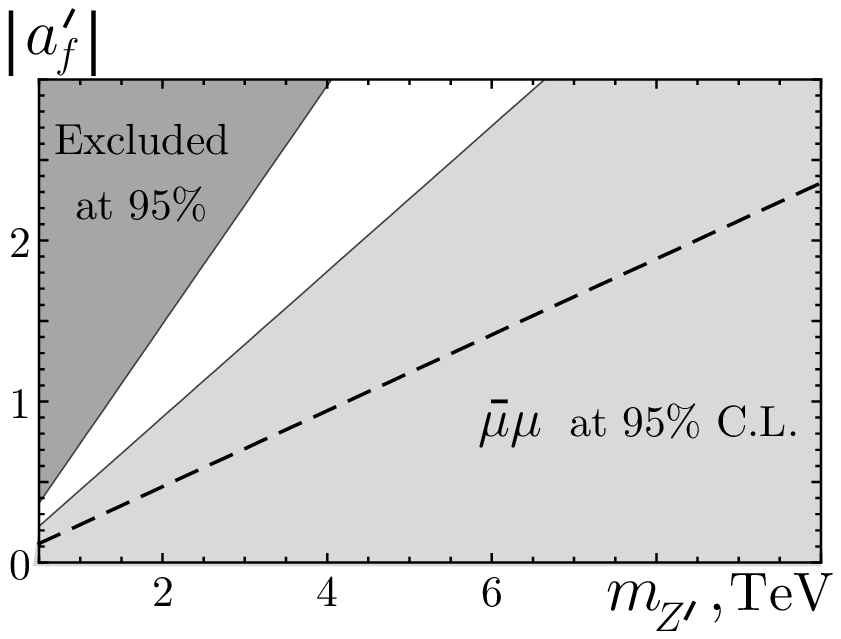,width=55mm}}
\caption{The maximum likelihood values and the confidence
intervals for the $Z$--$Z'$ mixing angle ($\theta_0$) and the
axial-vector couplings to the SM fermions ($a'_f$) by the LEP 2
data. The values excluded at 95\% CL by the many-parameter fit of
$e^+e^-\to l^+l^-$ are shown in dark gray. The results of fits
based on the one-parameter observables are shown in light gray for
$e^+e^-\to \mu^+\mu^-,\tau^+\tau^-$ (left) and for $e^+e^-\to
\mu^+\mu^-$ (right). The maximum likelihood value are plotted as
the dashed lines. The dotted lines correspond to the maximum
likelihood values obtained from the LEP1 data.}
\label{fig:fitFigs}
\end{figure}

\begin{figure}
\centerline{\psfig{file=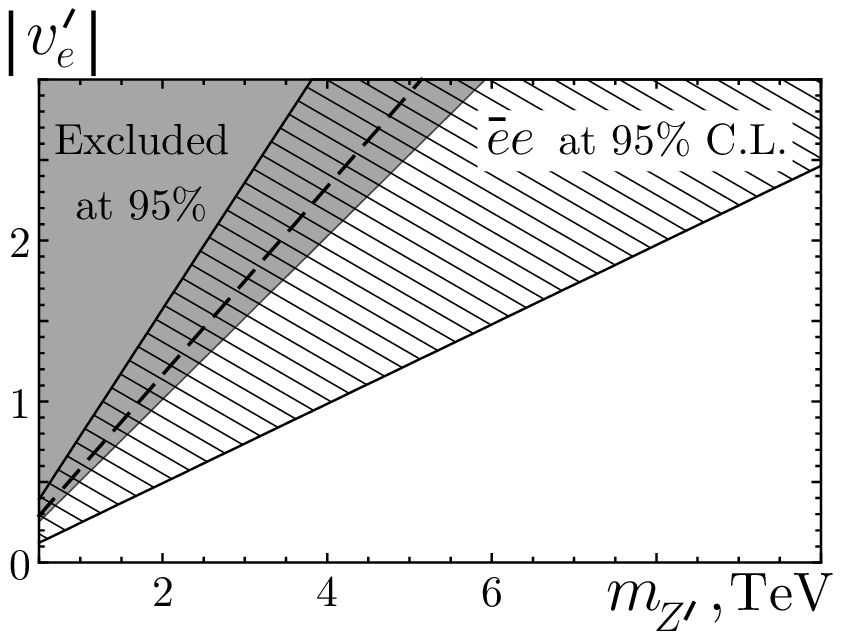,width=55mm}\hfil
\psfig{file=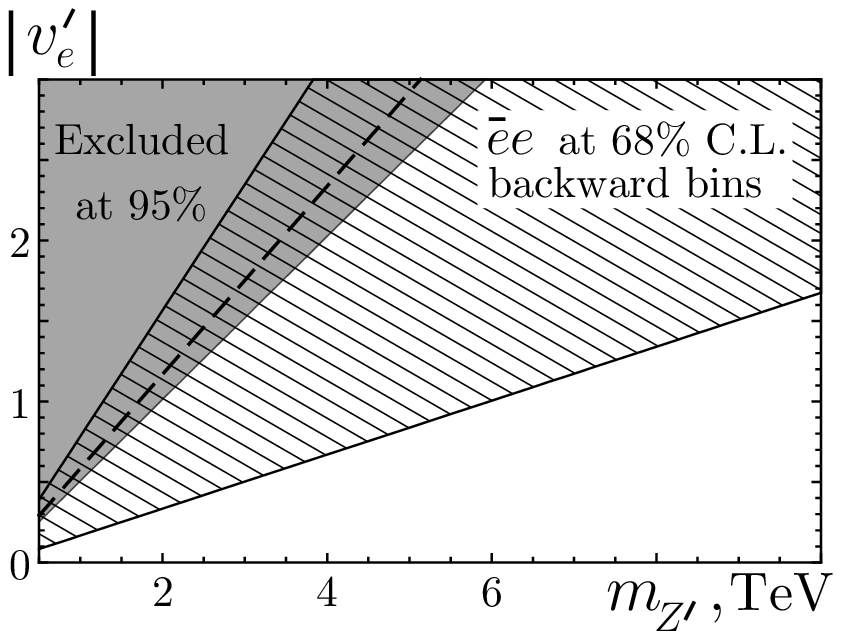,width=55mm}} \caption{The maximum
likelihood values and the confidence intervals for the vector
coupling to electron current ($v'_e$) by the LEP 2 data. The
values excluded at 95\% CL by the many-parameter fit of $e^+e^-\to
l^+l^-$ are shown in dark gray. The left panel represents the
results of fits based on the one-parameter observable for
$e^+e^-\to e^+e^-$. The right panel shows the 1$\sigma$ CL area
for the many-parameter fit of backward bins of $e^+e^-\to e^+e^-$.
The maximum likelihood value are plotted as the dashed lines.}
\label{fig:fitFigs1}
\end{figure}

In Table \ref{fitCLI} we present the confidence intervals for the
fitted parameters. It gives a possibility to estimate the
uncertainty of the $Z'$ couplings as well as the lower bounds on
the parameters. The results of fits are also shown in Figs.
\ref{fig:fitFigs}-\ref{fig:fitFigs1}. To summarize them we note
that the data of the LEP2 experiments are compatible at 1-2
$\sigma$ level with the existence of the not heavy $Z'$ boson.


Now we compare the above results with the ones of other fits
accounting for the $Z'$ presence. As it was mentioned in
Introduction, LEP collaborations have determined the model
dependent low bounds on the $Z'$ mass which vary in a wide energy
interval dependently on  a model. The same has also been done for
Tevatron experiments. The  modern low bound is $m_{Z'} \ge 850$
GeV. It is also well known that though almost all the present day
data are described by the SM
\cite{Alcaraz:2006mx,Abbiendi:2003dh,Abbiendi:1998ea,Ackerstaff:1997nf,Abdallah:2005ph,LEP1:2005ema},
the overall fit to the standard model is not very good. In Ref.
\cite{Ferroglia:2006mj} it was showed that the large difference in
$\sin^2 \theta^\mathrm{lept}_\mathrm{eff}$ from the
forward-backward asymmetry $A^b_{fb}$ of the bottom quarks and the
measurements from the SLAC SLD experiment can be explained for
physically reasonable Higgs boson mass if one allows for one or
more extra $U(1)$ fields, that is $Z'$. A specific model to
describe $Z'$ physics of interest was proposed which introduces
two type couplings to the hyper charge $Y$ and to the
baryon-minus-lepton number $B-L$. Within this model by using a
number of precision measurements from LEP1, LEP2, SLD and Tevatron
experiments the parameters $a_Y$ and $a_{B-L}$ of the model were
fitted. The presence of $Z'$ was not excluded at 68\% CL. The
value  of  $a_Y$ was estimated to be of the same order of
magnitude as in  our analysis and is comparable with values of
other parameters detected in the LEP experiments.  The erroneous
claim that $a_Y$ is two order less then the value derived from our
Table \ref{tab:sum:1} is, probably, a consequence of some missed
factors. The upper limit on the mass was also obtained $m_{Z'} \le
2.6 $ TeV .

These two analyzes are different but complementary.  A common
feature of them is an accounting for the $Z'$ gauge boson as a
necessary element of the data fits. The results are in accordance
at 68-95\% CL with the existence of the  $Z'$ which has a
good chance to be discovered at Tevatron and/or LHC.

\end{document}